\documentclass[superscriptaddress,pra,twocolumn,longbibliography]{revtex4-2}
\usepackage{amsmath,amssymb,amsthm}
\usepackage{easybmat,comment}
\usepackage[colorlinks=true,citecolor=blue,urlcolor=blue, linkcolor = magenta]{hyperref}
\usepackage[pdftex]{graphicx}
\usepackage{times,txfonts}
\usepackage{mathrsfs}
\usepackage{braket}
\usepackage{color}
\usepackage{natbib}
\usepackage{subcaption}
\usepackage{ragged2e}
\usepackage{commath}
\usepackage{mathtools}
\usepackage{amsmath}
\DeclareCaptionJustification{justified}{\justifying}
 \usepackage[compat=0.4]{yquant}
\newcommand{\be}{\begin{equation}}
	\newcommand{\ee}{\end{equation}}
\newcommand{\ba}{\begin{eqnarray}}
	\newcommand{\ea}{\end{eqnarray}}
\newcommand{\ketbra}[2]{|#1\rangle \langle #2|}
\renewcommand{\ketbra}[3]{|#1\rangle_{#2}\langle #3|}

\newcommand{\half}{\frac{1}{2}}

\usepackage{comment}
\usepackage{tikz}
\begin{document}
	
\title{Quantum steganographic protocols using degenerate and entanglement-assisted quantum codes}
\author{Sanjoy Dutta\textsuperscript{}}
   \affiliation{Poornaprajna Institute of Scientific Research (PPISR), Bidalur post, Devanahalli, Bengaluru 562164, India}
   \affiliation{Graduate Studies, Manipal Academy of Higher Education, Madhava Nagar,
Manipal 576104, India}
\author{Nihar Ranjan Dash\textsuperscript{}}
   \affiliation{Indian Institute of Technology Jodhpur, Rajasthan 342030, India}
\author{Subhashish Banerjee \textsuperscript{}}
   \affiliation{Indian Institute of Technology Jodhpur, Rajasthan 342030, India}
\author{R. Srikanth}
   \email{srik@ppisr.res.in}
   \affiliation{Poornaprajna Institute of Scientific Research, Bidalur Post, Devanahalli, Bengaluru 562164, India}




\begin{abstract}
Steganography is the art of concealing secret information by embedding it in an apparently innocent-looking message. Quantum steganography applies the principles of quantum mechanics to traditional steganography and, compared to the latter, offers significant advantages, including heightened security, improved concealment, and increased data-hiding capacity. Traditionally, quantum steganography disguises the covert communication as channel noise, which is corrected using preshared classical randomness. This method requires the steganalytic eavesdropper Eve to overestimate the level of channel noise, so that the bounds on the stego channel capacity depend on this assumed gap in Eve's knowledge of the channel. In this work, we point out that by means of preshared quantum entanglement the secret message can be encoded into nonlocal correlations, obviating the need for such an assumption of Eve's ignorance. Consequently, the capacity bounds on the stego channel can then come from the channel capacity of the quantum communication channel. We introduce three such entanglement-based quantum steganographic protocols that make use of catalytic quantum error-correcting codes (QECCs), degenerate entanglement-assisted QECCs, or the phase bit of preshared entanglement. Here catalytic QECCs enable recycling entanglement, while entanglement assistance allows both sender and receiver to contribute to the protocol's secrecy. We derive upper and lower bounds on the secrecy capacity of each protocol, and demonstrate their practical robustness.
\end{abstract}
\maketitle

\section{Introduction}\label{sec-intro}
Steganography is a technique to hide secret data by embedding the data within innocent-looking cover media, so that the existence of the hidden data remains undetectable to unintended recipients \cite{simmons1984prisoners, avritzer2024quantum}. The cover medium may be an audio signal \cite{dutta2020overview}, a digital image \cite{hu2020quantum}, a video file \cite{liu2021novel}, a cryptogram \cite{Qu2022Covert}, a text file, or headers of a network protocol. In the traditional setting, steganography can be motivated by considering Alice and Bob, who have been imprisoned in two geographically separated cells of a penitentiary.  The prison warden (Eve) allows them to communicate by swapping messages through a courier loyal to the warden. Under the circumstances, they have to exchange secret messages steganographically in order to avoid rousing the warden's suspicion.  Quantum steganography is an emerging field that applies principles of quantum information science to steganography. These principles, such as superposition, entanglement, and no-cloning, help to conceal information in a way that is theoretically more imperceptible, more secure against eavesdropping, and endowed with greater information capacity than classically possible \cite{qu2018novel}. As quantum technologies become increasingly ubiquitous in the near future and the quantum Internet of Things starts to take shape in smart cities, the demand for secure and covert communication methods is likely to surge, making quantum steganography an essential tool for safeguarding sensitive information in the emerging quantum era \cite{biswas2024advancing}. Here it is worth noting that while steganography is closely related to cryptography, indeed typically both requiring equal preshared resources \cite{sanguinetti2016perfectly}, yet the latter lacks the imperceptibility requirement central to the former \cite{taha2019combination}. 

The earliest quantum steganographic ideas can be traced to Terhal \textit{et al.} \cite{terhal2000hiding}, who introduced a technique to embed classical bits securely in entangled states. Gea-Banacloche \cite{gea2002hiding} proposed another early scheme for quantum steganography by embedding hidden information in the syndromes of a simple error-correcting code—the three-qubit repetition code. However, the resulting channel lacked stealth and had not been quantitatively analyzed. Subsequently, Shaw and Brun \cite{shaw2011quantum} proposed a steganographic scheme based on error-correcting codes and preshared classical bits, that ensured the cover traffic looked innocuous. Further, these authors analyzed the detectability and added performance metrics. Building on this idea, Ref. \cite{sutherland2019quantum} computes the maximum achievable rate of such covert communication if the eavesdropper (Eve) has imperfect knowledge of a noisy quantum channel’s parameters. Unlike classical or earlier quantum steganography methods where the cover data are modified to carry the secret, Mihara proposed hiding the secret in the nonlocal correlations of preshared entangled states, ensuring that the cover data themselves remain unaltered and indistinguishable from unsuspicious transmissions  \cite{mihara2015quantum}. Similarly, Qu \textit{et al.} \cite{qu2018novel} propose using preshared entanglement, in the form of Brown states rather than Bell pairs, optimizing for resistance to phase-flip noise and for stealth in secure direct communication. 

Recent studies in quantum steganography have explored newer methods of practical implementation, among them a twin-threshold scheme, where two predefined noise bounds are used to distinguish between cover data and stego data in a quantum communication channel \cite{tudorache2021quantum}; quantum image steganography encompassing quantum watermarking and quantum image encryption \cite{min2022quantum}; exploiting novel digital techniques for embedding the secret data in a cover image \cite{hu2020quantum}; a proposal to exploit quantum walk for steganographic image transfer in cloud systems \cite{abd2020secret}; the use of continuous-variables systems, where the secret bit is encoded in the position and momentum quadratures of coherent states \cite{joshi2022hide}; and an optical implementation in a Mach-Zehnder setup where the cover data are encoded in the polarization domain of photons, while the hidden information is embedded into the entanglement domain of the entangled photon pairs \cite{nagy2020quantum}. Mihara has proposed quantum steganography protocols for quantum networks by suitably adapting standard quantum key distribution channels \cite{mihara2017multi}. In a portent to an important future trend, machine-learning techniques have been integrated to improve the detection, embedding, and extraction of hidden information in quantum channels used for steganography \cite{Wang2021MLQuantum}.

The present work explores three approaches to efficient steganographic protocol design: the use of catalytic quantum codes, degenerate entanglement-assisted (EA) quantum error-correcting codes (QECCs), and the phase bit of encoded nonlocal correlations, none of which, to the best of our knowledge, has been applied in a stego context before. Accordingly, the rest of the manuscript is structured as follows. The main conceptual and mathematical preliminaries required are briefly introduced in Sec. \ref{sec-prelim}. The broad attack models of a staganalytic eavesdropper are discussed in Sec. \ref{sec:obviate}, where we also indicate corresponding counterstrategies of the sender and receiver for the class of entanglement-based protocols proposed in this work. We then introduce and investigate the performance of catalytic quantum codes in order to minimize the need for preshared ebits (Sec. \ref{sec:catalytic-QECC}). This involves utilizing an initial auxiliary resource (here an ebit, ``the catalyst'') to aid the encoding and decoding, but does not consume the catalyst or alter it irreversibly. 

A fundamental nonclassical feature of QECCs is error degeneracy, the feature whereby two distinct correctable errors are indistinguishable by syndrome measurement, and thus identically correctable \cite{smith2007degenerate}. Barring certain difficulty in decoding, degeneracy is in principle welcome for quantum communication because it implies that a larger number of errors can be tolerated, leading to improved code capacity and better error thresholds \cite{chiribella2011quantum}. By contrast, in the context of quantum stego protocols where noise operators code for secret bits, degeneracy can be undesirable as it creates redundancies in the error space. In Sec. \ref{sec:degenerate-QECC} we show that the quantum degeneracy of errors can be steganographically exploited, with the aid of EA QECCs, to design protocols where both sender and receiver of stego messages contribute to the secrecy and security. EA QECCs are quantum codes that protect information by using both local encoding and preshared entanglement (ebits) between sender and receiver, enabling correction of errors even when the usual stabilizer constraints do not hold. 

Mihara's protocol \cite{mihara2015quantum} made use of prior entanglement and nonlocal correlations as the basis of steganographic communication, deviating from the theme of encoding stego bits in channel errors. As such, the encoding scheme employed there exploits the phase bit of the encoded entangled state. In Sec. \ref{sec:scheme-3} we explore the use of alternate forms of embedding the secret in the correlations, and the corresponding challenges to the security, secrecy, and decoding mechanism. Finally, we present our conclusions and discussions in Sec. \ref{sec:conclusion}, outlining various future directions indicated by our work.

\section{Preliminaries}\label{sec-prelim}
\color{black}
\subsection{EA QECC} 
An EA QECC uses preshared entanglement to improve the noise resistance of a QECC. Given ebits $\big(\ket{\Phi^+}_{AB}^{\otimes e}\big)$ between the two communicating parties, an EA QECC can be constructed from any linear classical code by relaxing the constraint of dual containing classical code, which is necessary for the construction of a stabilizer QECC \cite{hsieh2007general, pereira2022entanglement}, defined by a set of commuting stabilizer generators. More generally, quadratic constraints are imposed on the classical code to obtain a quantum code of qubits or qudits. These constraints take the form of dual containment (in the case of stabilizer or CSS codes) or symplectic self-orthogonality (in the case of general qudit
stabilizer codes), and ensure that the resulting error correcting code enforces a commuting relationship among the stabilizers.  

In the case of an $[[n,k,d;e]]$ EA QECC, the stabilizer group over the $n$ qubits can be non-Abelian, thus allowing us to relax the above quadratic constraints on the classical linear codes. Here Alice encrypts the $k$-qubit state $\ket{\psi}$ in $n$ qubits, and appending the $e$ extra Pauli operators, the stabilizers are rendered commuting. During encoding, Alice appends $a \equiv n-k-e$ ancillae before an encryption operation $(U_A \otimes I_B)$ on her $n$ particles to produce the encoded state $ (U_A \otimes I_B)\ket{\psi} \otimes\ket{0}^{a} \otimes \ket{\Phi^+}_{AB}^{\otimes e}$ \cite{lidar2013quantum}. 
 
\subsection{Catalytic QECC}
In the context of quantum communication using EA QECC,  Alice consumes preshared ebits prepared in the state $\ket{\Phi^+}$.  Assume that they have a small number of preshared ebits but she has the resource of locally available ebits.  To compensate for the loss of ebits during communication, Alice includes $e$ halves of the local ebits into the state $\ket{\psi}$ to be encoded, thus $k\ge e$. After her transmission of the encoded register to Bob, and his decoding, the entangled state $\ket{\Phi^+}^{\otimes e}$ will be established as the preshared entanglement for the next cycle. This technique of constructing catalytic QECC can also be applied to standard QECCs encoding entangled states. \cite{brun2006correcting,brun2014catalytic}.

\subsection{Degenerate QECC}
Degeneracy of errors is the phenomenon whereby distinct error operators from the correctable set $\mathcal{E}$ of errors act on a quantum code to send it to the same erroneous state \cite{smith2007degenerate}.  By way of example, consider the three-qubit code: 
\begin{align}
   \ket{w=0_L} &= \frac{1}{\sqrt{2}}(\ket{000} + \ket{111})\nonumber\\
    \ket{w=1_L} &= \frac{1}{\sqrt{2}}(\ket{010} + \ket{101}.
\end{align}
The code space spanned by $\ket{0_L}$ and $\ket{1_L}$ is driven to the same erroneous subspace by the operators $Z_1$ and $Z_3$.

\subsection{Bounds on entanglement-assisted channel capacity}\label{subsec:bounds_on_ea_channel_capacity}
The entanglement-assisted classical capacity of a channel, $\mathcal{E}$, is given by \cite{bowen2002entanglement}
\begin{align}
    C_{E}=\underset{\rho}{\rm max}[S(\rho)+S(\mathcal{E\rho)}-S\big((\mathbb{I}\otimes \mathcal{E})\ket{\rho}\bra{\rho}\big)],
    \label{eq:eaclasscialcapacity}
\end{align}
where $\ket{\rho}$ is any purification of the state $\rho$ and $S(\rho)$ is the von Neumann entropy of the state $\rho$, given by $S(\rho)=\rm{Tr}\rho \log_{2}\rho$. By considering the above result in conjunction with quantum dense coding, it can be shown that the entanglement-assisted quantum-information capacity is given by
\begin{equation}
    Q_E=\frac{1}{2}C_E.
    \label{eq:half}
\end{equation}
Unlike classical information, even with preshared entanglement, only $m$ qubits of information can be transferred by transmitting $m$ physical qubits.

When the quantum information is encoded into EA QECC, the Hamming and Gilbert-Varshamov (GV) bounds can be used to obtain the upper and lower bounds on the quantum capacity. Taking log on both sides of the (nondegenerate) Hamming bound $\sum_{j=0}^t 3^j{n \choose j} \le 2^{n+c-k}$, and applying the Stirling approximation, we obtain
\begin{equation}
    \frac{k}{n} \le 1+\frac{c}{n} - \frac{t}{n}\log(3) - H\left(\frac{t}{n}\right),
\label{eq:uppercapacity}    
\end{equation}
where $t$ is the maximum number of correctable errors, log is to base 2, and $H$ is the binary Shannon entropy. 
Analogously the GV bound for good QECC codes $\sum_{j=0}^{2t} 3^j{n \choose j} \ge 2^{n+c-k}$ yields
\begin{equation}
    \frac{k}{n} \ge 1+\frac{c}{n} - \frac{2t}{n}\log(3) - H\left(\frac{2t}{n}\right),
\label{eq:lowercapacity}    
\end{equation}
where $t$ is the maximum number of correctable errors, log is to base 2, and $H$ is the binary Shannon entropy.

\section{Using entanglement to hide secrets: attack models and security \label{sec:obviate}}
Steganography attack models, largely categorized under steganalysis, are defined by the amount of information available to the attacker about the cover medium (the carrier) and the stego object (the carrier with hidden data). These models aim to either detect, extract, or destroy hidden data. In the quantum context, steganalysis aims to attack specific target quantum states and protocols. Broadly, the steganalyst Eve seeks strategies to detect a concealed covert communication between Alice and Bob, and possibly jam it. A passive steganalytic method would be to monitor the communication channel to detect covert signals; an active method of intervention would be to intercept and measure the particles transmitted from Alice to Bob. Thus, the main requirement of a stego protocol is imperceptibility, or secrecy, whereby Alice can transmit secret information hidden in innocent-looking data. Apart from this, we require a robust secrecy capacity $
    C_s \equiv \max(C_{\rm legitimate} - C_{\rm eavesdropper}, 0),$
where the right-hand side describes the difference between the capacity of the legitimate channel between Alice and Bob and the capacity of the eavesdropper's channel. Finally, we require security, to ensure the protection of the secret if detected. 

In quantum stego protocols where Alice and Bob deliberately disguise messages as channel noise, two requirements arise, which bound the quantum stego capacity \cite{sutherland2019quantum}: (a) an assumption that Eve overestimates the noise level, i.e., there is a gap between the true channel noise level, given by parameter $p$, and Eve's perception thereof, assumed to be a larger level, $p+\delta p$; and (b) fine-tuning the message so that it can be passed off as noise that fits this ignorance gap $\delta p$.

Suppose the level of channel noise $\mathcal{N}_p$ in a \textit{bit-flip or dephasing channel} is $p$ (probability to flip a bit or the phase $\pm1$), but Eve has been deceived systematically by Alice and Bob to believe that the noise is described by $\mathcal{N}_{p+\delta p}$. For sufficiently large $N$, the total number of typical strings is about $2^{Nh(p+\delta p)}$, whereas from Alice's perspective, her any input error pattern corresponds to a volume of typical strings of $2^{Nh(p)}$. Thus an upper bound on the number of retrievable secret messages from Alice to Bob is given by rate $R$ such that $
    2^{NR} \equiv 2^{Nh(p+\delta p) - Nh(p)}.$ 
It is straightforward to show that given $\mathcal{N}_{p+\delta p} = \mathcal{N}_{p} \circ \mathcal{N}_{q}$, where $q$ is the noise level corresponding to Alice's covert communication, then $q \equiv \delta p/(1-2p)$.

In contrast to the above, all the protocols proposed here will be based on preshared entanglement, which is used to encode the secret data via nonlocal correlations, rather than error syndromes. This exploits a quintessential quantum feature whereby identical mixed states can be purified into distinct states in a larger state space. Quite generally, suppose the states that encode secret bit $b=0$ and $1$ are denoted by
\begin{equation}
    \ket{\chi(b)}_{AB} = \sum_j \ket{\lambda_j^{(b)}}_A\ket{\mu_j}_B,
\end{equation}
such that (a) $\ket{\chi(b)}_{AB}$ can be encoded by Alice's local action alone starting from a fixed initial state $\ket{\chi(*)}_{AB}$ and, (b) furthermore
\begin{equation}
 {\rm Tr}_B\ketbra{\chi(*)}{AB}{\chi(*)} =   {\rm Tr}_B\ketbra{\chi(0)}{AB}{\chi(0)} = {\rm Tr}_B \ketbra{\chi(1)}{AB}{\chi(1)}.
    \label{eq:equal}
\end{equation}
This ensures that the reduced state transmitted by Alice carries no information of the encoded bit or even whether a secret bit is encoded. The action of any noise superoperator $\mathcal{E}$ modifies each of the three reduced states in Eq. (\ref{eq:equal}) identically, and thus gives no advantage for Eve's detection. Another important consequence is that the channel capacity or rates (introduced in Sec. \ref{subsec:bounds_on_ea_channel_capacity}) or a fixed factor thereof can serve as the secrecy capacity or rates, under suitable conditions. This is discussed in the case of each protocol later below.

Suppose Eve makes an active intervention to extract information. Below we show that given the states are quantum encoded, any intervention that produces a correctable error leads to no information gain for Eve. Let $\ket{\psi_L}_{AB}$ denote the logical state in an EA QECC with code space projector $P$. A probe (Eve's or the environment) initially in state $\ket{e_0}_E$ interacts with the system via a unitary $U_{AB}$, producing an entangled state:
$$
(U_{AE} \otimes \mathbb{I}_B) \big( \ket{\psi_L}_{AB} \otimes \ket{e_0}_E \big) = \sum_j K_j \ket{\psi_L}_{AB} \otimes \ket{e_j}_E,
$$
where $\{K_j\}$ are correctable errors satisfying the Knill-Laflamme condition $P K_i^\dagger K_j P = \alpha_{ij} P$ for some Hermitian matrix $\alpha$.

It follows that the reduced state of the probe is given by
$$ 
\sum_{k,l} {_{AB}}\bra{\psi_L}K^{\dagger}_l K_k\ket{\psi_L}_{AB}\ketbra{e_k}{E}{e_l} = \sum_{k,l} \alpha_{l,k} \ketbra{e_k}{E}{e_l},
$$
implying that the probe is independent of the logical state. Thus error correctability ensures that the probe interacts with and gains information only of the syndrome subsystem, but not the encoded state. If the error correction conditions are violated, then the expectation values $\bra{\psi_L}K^{\dagger}_l K_k\ket{\psi_L} \ne \alpha_{l,k}$, and the probe state $\rho_E$ may depend on $\ket{\psi_L}$. However, here the inherent security aspect of the protocol comes into play, leading to an information vs disturbance tradeoff.

\section{Protocol 1: Using Catalytic QECC}\label{sec:catalytic-QECC}
Sufficient quantity of said resource must be preshared as it is consumed during each round of the stego protocol. However, distributing entanglement, especially over noisy channels, can be technologically challenging to the secrecy requirement, in contrast to the security requirement of quantum cryptography.  While the classical resource here is relatively cheap, albeit potentially insecure, by contrast the quantum resources can be secure, but costly to pre-share.
Our first protocol proposes that a catalytic QECC-- rather than conventional QECC-- should be used for noise encoding the secret information. With the catalysis, a low number of initially preshared ebits will suffice, since part of it will be recycled to replace the consumed ebits. 

The protocol employs a dense-coding-like scheme, in which entanglement is replenished after each round by using a QECC of sufficiently large rate and an initial stock of local entanglement.
\subsection{Protocol 1}
The following steps are sequentially executed.
\begin{enumerate} 
    \item Alice and Bob pre-share an ebit $\ket{\Phi^+}_{AB}$, and agree on a $[[n,k,d; 0]]$ QECC, with $k\ge2$. (More generally, if the secret rate is $k_s$, then $k = 2k_s$.) Further, Alice prepares a local ebit $\ket{\Phi^+}_{l_1,l_2}$.  
    \item Alice decides on two bits-- the cover message $w$ and the secret bit $b$, and employs dense coding to prepare the Bell state $\ket{\eta(w,b)}_{AB} \equiv (\ket{0,w} + (-1)^b\ket{1,\overline{w}})_{AB}$, by applying the local operation of $(X^{w}Z^{b})_A\otimes\mathbb{I}_B$. 
    \item Alice then encodes one half of her local ebit and her half of the ebit shared with Bob, i.e., the first two particles in right-hand side, 
    \begin{align}
        &\ket{\Phi^+}_{l_1,l_2}\ket{\eta(w,b)}_{AB} =  \half\big(\ket{0,0}_{l_1,A}\ket{0,w}_{l_2,B} \nonumber + \\ &(-1)^b\ket{0,1}_{l_1,A}\ket{0,\overline{w}}_{l_2,B} +  \ket{1,0}_{l_1,A}\ket{1,w}_{l_2,B} + \nonumber \\ &(-1)^b\ket{1,1}_{l_1,A}\ket{1,\overline{w}}_{l_2,B}\big) \nonumber 
    \end{align} 
    obtaining
     \begin{align}
        & \half\big(\ket{(0,0)_L}_{l_1,A}\ket{0,w}_{l_2,B} +(-1)^b\ket{(0,1)_L}_{l_1,A}\ket{0,\overline{w}}_{l_2,B} + \nonumber \\
        & \ket{(1,0)_L}_{l_1,A}\ket{1,w}_{l_2,B} +(-1)^b\ket{(1,1)_L}_{l_1,A}\ket{1,\overline{w}}_{l_2,B}\big)
        \label{eq:later}
   \end{align} 

        \item She transmits her particles to Bob over a noisy channel. After performing the necessary quantum error correction using her and his particles jointly, and decoding the resultant state, he obtains 
        $ 
        \ket{\Phi^+}_{l_1,l_2}\ket{\eta(w,b)}_{A,B},
        $
        where particles $A, B, l_1$ are now with Bob. 
        \item On the particles $A,B$ Bob applies a controlled-NOT (CNOT) with the control on $B$ followed by a Hadamard on $B$. This results in the transformation
        $$ \ket{\eta(w,b)} \longrightarrow \ket{w}\ket{b}.$$
        \item The state $\ket{\Phi^+}_{l_1,l_2}$ will serve as the shared ebit of the next round.
    \end{enumerate}
The catalytic aspect, which is the replenishment of the entanglement consumed, ensures that the initial preshared ebit suffices to transmit any number of secret qubits, over subsequent rounds. Note that because we employ a dense-coding protocol, Alice is restricted to transmitting a classical cover message and classical secret message. Below we describe a probabilistic protocol for transmitting secret qubits.

Consider an arbitrary single-qubit state $\cos(\alpha)\ket{b=0} + \sin(\alpha)e^{i\beta}\ket{b=1}$ as a secret, as well as an arbitrary cover message $\cos(\mu)\ket{w=0} + \sin(\mu)e^{i\nu}\ket{w=1}$. Alice by means of her local operations must prepare the state 
\begin{align}
    \ket{\eta(\textbf{b,w})} &= \cos(\mu)\big(\cos(\alpha)\ket{\Phi^+}+\sin(\alpha)e^{i\beta}\ket{\Phi^-}\big) \nonumber \\ &+ \sin(\mu)e^{i\nu}\big(\cos(\alpha)\ket{\Psi^+}+\sin(\alpha)e^{i\beta}\ket{\Psi^-}\big).
    \label{eq:DCqubit}
\end{align}
 To do so, Alice prepares the secret state and cover message in qubit ancillae, and applies a C-PHASE and CNOT gate with the former and the latter as the respective control qubits, while the target qubit is her half of the entanglement shared with Bob. This yields
\begin{align*}
&\big(\cos(\mu)\ket{0} + \sin(\mu)e^{i\nu}\ket{1}\big) \big(\cos(\alpha)\ket{0} + \sin(\alpha)e^{i\beta}\ket{1}\big) \ket{\Phi^+}\\ &\rightarrow  \ket{\eta(\textbf{b,w})}.
 \end{align*}
Measuring  the first two registers in the diagonal ($XX$) basis, she produces the required state $\ket{\eta(\textbf{b,w})}$ with probability $\frac{1}{4}$, conditioned on obtaining $\ket{+,+}$. It is assumed that Alice can publicly communicate to Bob the instance of successful encoding, without arousing Eve's suspicion. 

\subsection{Secrecy, security, and rate bounds}
It follows from Eq. (\ref{eq:later}) that the state of the two registers transmitted by Alice to Bob is
\begin{align}
    \rho_{l_1,A} &= \frac{1}{2}\big(\ketbra{(00)_L}{l_1 A}{(00)_L} + \ketbra{(01)_L}{l_1 A}{(01)_L} \nonumber \\
    &+ \ketbra{(10)_L}{l_1 A}{(10)_L} + \ketbra{(11)_L}{l_1 A}{(11)_L}\big), 
    \label{eq:secrecy1}
\end{align}
i.e., the maximally mixed state in the two-logical-qubits code space, irrespective of $w, b$, satisfying the security requirement Eq. (\ref{eq:equal}). Under any channel noise described by Kraus operators $\{\mathfrak{E}_j\}$, the transmitted system's state transforms as $\rho_{l_1,A} \rightarrow \sum_j\mathfrak{E}_j \rho_{l_1,A} \mathfrak{E}_j^{\dagger}$, but remains independent of the secret bit $b$. Thus, the upper and lower bounds on steganographic rates come directly from the noisy channel capacity.

By way of illustration, we consider two well-known noise channels, the depolarizing and dephasing channels.  For a depolarizing channel given by $\rho \mapsto (1-p)\rho + \frac{p}{3}(\mathcal{X}+\mathcal{Y}+\mathcal{Z})(\rho)$ (where $\mathcal{X}, \mathcal{Y}, \mathcal{Z}$ are the Pauli superoperators, and $p \in [0,\frac{3}{4}]$), from Eq. (\ref{eq:eaclasscialcapacity}), one finds that the classical capacity is given $C_E = 2 - p\log(3) - H\left(p\right)$. It follows that the secrecy capacity is half of this (noting that only bit $b$, and not $w$, serves as secret): 
\begin{equation}
    R_s^C \le 1 - \frac{p}{2}\log(3) - \frac{1}{2}H\left(p\right),
\label{eq:capdepol0}
\end{equation}
where the superscript $C$ indicates that the transmitted secret is classical. Writing $R_s^C \equiv \frac{k_s^C}{n}$, we see that the number of stego bits transmitted scales linearly with $n$.  

We note that the upper bound on the secret rate for this channel and the quantum capacity coincide. It is then reasonable to expect that Eq. (\ref{eq:lowercapacity}), with $c:=0$, allows us to lower bound the rate applicable to this steganographic protocol:
\begin{equation}
    R_s^C \ge 1 - 2p\log(3) - H\left(2p\right).
    \label{eq:floordepol0}
\end{equation}
Note that although the protocol involves shared entanglement, this is not used as entanglement assistance for the quantum code, and hence $c=0$. From Eqs. (\ref{eq:capdepol0}) and (\ref{eq:floordepol0}), we find that the upper and lower bounds on $R_s^C$ vanish for $p=\frac{3}{4}$ and $p\approx 0.095$, respectively. This entails that no secret transmission is possible above 75\% noise level whereas nonzero secret transmission is guaranteed (asymptotically) if the noise level is below $\sim9\%$.

For a dephasing channel given by $\rho \mapsto (1-p)\rho + p\mathcal{Z}(\rho)$, the entanglement-assisted classical capacity is $C_E=2-H(p)$. Note that this does not vanish for maximal noise $(p=\frac{1}{2})$ because the channel leaves the populations unchanged as in a noiseless classical channel (even though it fully destroys the coherences). For this protocol,
\begin{equation}
    R_s^C \le 1 - \frac{1}{2}H(p),
    \label{eq:capdephas}
\end{equation}
which, again, coincides with capacity $Q_E$. As before, Eq. (\ref{eq:lowercapacity}) is used to lower bound the secret rate. For large $n$ and assuming the dephasing channel (which has only one rather than three types of errors), we can use the asymptotic quantum Gilbert-Varshamov bound to estimate the allowed quantum code rate, by setting $3 \rightarrow 1$ in Eq. (\ref{eq:lowercapacity}). This then asserts that there exist good $[[n,k,d]]$ codes, i.e., a code with rate
$
  \frac{k}{n} \geq 1- H_2(\delta). 
$
Therefore, the asymptotic secrecy rate $R_s^C$ for our catalytic stego protocol is lower bounded by
\begin{align}
  R_s^C \geq 1 - H(2p), 
  \label{eq:stegoGV}
\end{align}
where asymptotically $d = 2t$ and we set $\frac{t}{n} := p$. Equations. (\ref{eq:capdephas}) and (\ref{eq:stegoGV}) both entail a linear scaling of secret bits with $n$. 

From Eq. (\ref{eq:capdephas}), we find that the upper bound is nonvanishing for all $p$, attaining its minimum of $\frac{1}{2}$ at maximal dephasing, $p=\frac{1}{2}$. This is because encoding in computational basis makes the information impervious to phase noise. On the other hand, from Eq. (\ref{eq:stegoGV}), we find that the lower bound on $R_s^C$ vanishes for $p=\frac{1}{4}$. This shows that this protocol is quite robust in the presence of dephasing noise, more so than in the presence of depolarizing noise.

\section{Using degenerate entanglement-assisted QECC}
\label{sec:degenerate-QECC}
When secret information is camouflaged as noise in a stego protocol, the use of error degeneracy of the code is not expected to be helpful since it would create either ambiguity (if each error encodes a distinct information) or superfluity (if degenerate errors encode the same information). Here we will find that degeneracy can be useful, provided QECCs are replaced with EA QECCs. The use of EA QECC in the following protocol contrasts with the preceding protocol or that of Ref. \cite{mihara2015quantum}, where preshared ebits are used in a way that is structurally independent of the QECCs used. 

\subsection{Protocol 2}
In the present protocol (which we shall call scheme $\mathfrak{Q}$), all preshared ebits are used to encode for the EA QECC. As in Protocol 1, the secret is encoded via nonlocal correlations rather than syndromes, allowing for the state with Alice to be uncorrelated with the encoded message. Syndromes will serve as keys to unlock the true secret message from the cover message, which is encoded as a conventional encoded quantum state. Unlike in protocol 1, here EA QECC encoding rather than dense coding is the method to send messages, enabling the transmission of quantum secrets deterministically.  

Alice uses a dual quantum-classical channel of transmission. She quantum encrypts the secret state $\ket{\varphi}$ and transmits the encoded state over the EA QECC channel. The classical key to decrypt the decoded state is encoded as an error syndrome in the transmission. In the ideal case of an underlying noiseless channel (as known to Alice and Bob but not to Eve), Bob decodes the received state and applies the Pauli correction as specified by the syndrome. In the noisy case, Bob still decodes the received state by the conventional method. If the channel is characterized by noise at level $p$, then Alice and Bob agree to implement an EA QECC that corrects noise up to level $p + \delta p$, where $\delta p$ is chosen according to the noise level $q$ by which Alice introduces noise corresponding to the encryption information. As we saw earlier, in case of a dephasing or bit-flip channel, $q = \delta p/(1-2p)$ (Sec. \ref{sec:obviate}).

A basic requirement for the present protocol is, as with Protocol 1, to meet the security requirement Eq. (\ref{eq:equal}): the reduced state with Alice must be independent of the encoded state or injected error. This can constrain the EA QECC and/or the qubits chosen to be with Bob. This condition is satisfied in the case of $[[n,k,d]]$ codes that have the \textit{quantum secret sharing} property \cite{cleve1999share}. By the stabilizer structure, the code can correct $(d-1)$ erasures, meaning that the remaining $n-d+1$ qubits suffice to reconstruct the state, knowing the coordinates of the erased qubits. Then, the no-cloning bound on secret sharing requires that $n-d+1>\frac{n}{2}$ (otherwise, the encoded state could be cloned), which entails
\begin{equation}
    n > 2d-2.
    \label{eq:nocloning}
\end{equation}
Imposing the secret sharing condition, with any $n-d$ or fewer qubits, no information about the encoded state is revealed. 
We can thus steganographically use a $[[n-d,k,d; d]]$ EA QECC, where we require that the number of qubits with Alice must exceed that of the shared entangled pairs, i.e., $n-d>d$, or $n>2d$. This is consistent with the no-cloning bound Eq. (\ref{eq:nocloning}), but stronger. Note that the Shor $[[9,1,3]]$ code satisfies this, whereas the five-qubit code $[[5,1,3]]$ does not.

Suppose the set of errors that Alice (resp., Bob) can steganographically apply to her (resp., his) qubits is denoted $\mathfrak{E}_A \equiv \{\mathfrak{e}_A\}$ (resp., $\mathfrak{E}_B \equiv \{\mathfrak{e}_B\}$). Importantly, as with protocol 1, the preshared entanglement in Alice's errors must be correctable and characteristic of the channel noise that Eve expects. The combination of the two legitimate parties' errors need not be correctable but must satisfy
\begin{equation}
    \mathfrak{e}_B \mathfrak{e}_A \in \mathcal{S} \cup (\mathfrak{E}_A-I) \circ \mathcal{N}(\mathcal{S}),
\label{eq:normalizer}
\end{equation}
where $\mathcal{N}(\mathcal{S})$ denotes the normalizer of the stabilizer $\mathcal{S}$. The basic idea here is that Alice encrypts her secret message into an innocent cover object, and encodes the decryption key via errors, which Bob decodes by syndrome measurement. In the event of being challenged by Eve, Bob applies an element randomly drawn from set $\mathfrak{E}_B$. The effectiveness of the method is clarified below.

\subsection{Secrecy, security, and rate bounds}
The key idea about using degeneracy is that, in the event of being challenged by Eve, Bob can neutralize Alice's message by jamming her key $\mathfrak{e}_A$ via his local application of error $\mathfrak{e}_B$. If Eq. (\ref{eq:normalizer}) is satisfied, Bob succeeds in eliminating or randomizing Alice's error. To achieve innocence (asymptotically), the entropy $H(q)$ in Alice's alphabet must be sufficiently low (as noted above).
To show this, suppose Alice encodes her (encrypted) cover message $\ket{\psi}_A$ into an EA code logical state $\ket{\psi_L}_{AB}$. Further, let her secret key bit correspond to error $\mathfrak{e}_A \in \mathfrak{E}_A$. She prepares the state $\mathfrak{e}_A\ket{\psi_L}_{AB}$ and transmits her qubits to Bob. As the errors $\mathfrak{e}_A$ are at most with noise level $p+\delta p$ and thus correctable, Bob obtains the secret key required to decrypt the cover message (normal mode). However, if Eve challenges the duo (challenge mode), Bob surrenders his qubits after applying to his qubits' random errors $\mathfrak{e}_B$ sampled over set $\mathfrak{E}_B$ by a probability distribution $\mathscr{P}$. Two possibilities arise:
\begin{description}
    \item[$\mathfrak{e}_B\mathfrak{e}_A \in \mathcal{S}$]: The two errors neutralize by virtue of degeneracy of $\mathfrak{e}_A$ and $\mathfrak{e}_B$. Eve interprets this as a possible action of the expected channel noise. 
    \item[$\mathfrak{e}_B\mathfrak{e}_A \in (\mathfrak{E}_A-I)\circ \mathcal{N}(\mathcal{S})$]: The logical state is rotated within the code space up to an allowed error. Eve interprets this as a possible action of the expected channel noise on a rotated state \footnote{In the second possibility above, the reason that we exclude the case $I\circ\mathcal{N}(\mathcal{S})$ is that the weight of an arbitrary error $\mathfrak{e}$ satisfies $\vert \mathfrak{e} \vert \le \lfloor \frac{d-1}{2}\rfloor$. Thus $\vert \mathfrak{e}_A\mathfrak{e}_B\vert \le d-1$, whereas any logical operator $O_L$ is such that $|O_L| \ge d$.}. 
\end{description} 
In either case, Eve only accesses the cover message and has no reason to suspect that the errors are keys to decrypt the true message.

For a depolarizing channel, from Eqs. (\ref{eq:eaclasscialcapacity}) and (\ref{eq:half}), we obtain an upper bound on the qubits that can be transmitted when shared entanglement is available: 
\begin{equation}
    R_s^Q \le 1  - \frac{p}{2}\log(3) - \frac{1}{2} H(p),
    \label{eq:capdepol}
\end{equation}
where the superscript $Q$ indicates that the transmitted secret is quantum rather than classical. For arriving at a lower bound, we consider the EA GV bound Eq. (\ref{eq:lowercapacity}), and the code $[[n,k,d\equiv 2t+1; c]]$ is chosen such that $\frac{t}{n} \le p$. This yields
\begin{equation}
    R_s^Q \ge 1+\frac{c}{n} - 2p\log(3) - H\left(2p\right),
\label{eq:lowercapacity2}    
\end{equation}
For a maximal entanglement code, characterized by $c=n-k$, the above yields
\begin{equation}
    R_s^Q \ge 1 - p\log(3) - \frac{1}{2}H\left(2p\right),
\label{eq:lowercapacity1}    
\end{equation}
which provides a lower bound for our protocol in the context of a depolarizing channel.

For a dephasing channel given by $\rho \mapsto (1-p)\rho + p\mathcal{Z}(\rho)$, Eqs. (\ref{eq:eaclasscialcapacity}) and (\ref{eq:half}) entail
\begin{equation}
    R_s^Q \le 1 - \frac{1}{2}H(p).
    \label{eq:capdeph1}
\end{equation}
For the lower bound, we consider the EA GV bound Eq. (\ref{eq:lowercapacity1}) adapted for a single type of Pauli error:
\begin{equation}
    R_s^Q \ge 1+\frac{c}{n} - H\left(2p\right).
\label{eq:lowercapacity2}    
\end{equation}
Here again, we can set $c:=n-k$ corresponding to maximal entanglement, which yields the most optimistic lower bound:
\begin{equation}
R_s^Q \ge 1 - \frac{1}{2}H\left(2p\right).
\label{eq:lowercapacity3}    
\end{equation}
The above rates allow us to assess the protocol's robustness. From Eqs. (\ref{eq:capdepol}) and (\ref{eq:lowercapacity1}), we find that the upper and lower bounds on $R_s^Q$ vanish for $p=\frac{3}{4}$ and $\frac{3}{8}$, respectively. This entails, as with protocol  1, that no secret transmission is possible above 75\% noise level. However, nonzero secret transmission is guaranteed (asymptotically) if the noise level is below 37.5\%. Protocol 2 then is more robust towards noise than protocol  1 for this type of channel (Fig. \ref{fig:depol}). Moreover, protocol 2 allows deterministic transmission of quantum secrets, whereas it is probabilistic in the case of protocol 1. 

In the case of the dephasing noise, from Eq. (\ref{eq:lowercapacity3}), we find that the lower bound also does not vanish. This reflects the fact that in certain classes of states, namely a mixture of the computational basis states, the mixture remains invariant under this type of noise.
\color{black}

\subsection{Example}
We now present a simple example of an EA QECC exploiting quantum degeneracy in the presence of a dephasing channel $\mathcal{E} \equiv (1-p) \mathcal{I} + p\mathcal{Z}$ $(0\le r\le1)$. Alice and Bob replace the channel with a noiseless one. The noiselessness need not be assumed (as noted above), but we only make the assumption for simplifying this example. We use the $[[6,1,3;3]]$ EA QECC constructed from the $[[9,1,3]]$ Shor code, represented by the  stabilizer generator set
\begin{align}
    \mathcal{S}=\{s_1=Z_1Z_2,s_2=Z_2Z_3,s_3=Z_4Z_5,s_4=Z_5Z_6,s_5=Z_7Z_8,\nonumber \\ s_6=Z_8Z_9,s_7=X_1X_2X_3X_4X_5X_6,s_8=X_4X_5X_6X_7X_8X_9\}.
\end{align}
The secret is encoded via the codewords
\begin{align}
    \ket{0_L} &= \frac{1}{\sqrt{2}} (\ket{00\textbf{0}} + \ket{11\textbf{1}}) (\ket{00\textbf{0}} + \ket{11\textbf{1}}) (\ket{00\textbf{0}} + \ket{11\textbf{1}}),\nonumber\\
    \ket{1_L} &= \frac{1}{\sqrt{2}} (\ket{00\textbf{0}} - \ket{11\textbf{1}}) (\ket{00\textbf{0}} - \ket{11\textbf{1}}) (\ket{00\textbf{0}} - \ket{11\textbf{1}}),
\end{align}
where the qubits $3, 6$, and $9$ (represented by the bold letter) are with Bob, while the other six are with Alice. Suppose the secret is the state $\ket{\varphi}$. The cover message will be the logical encoding of one of the quantum encrypted states: $\ket{\varphi} = \alpha\ket{0}+\beta\ket{1}, X\ket{\varphi}, Y\ket{\varphi}$, or $Z\ket{\varphi}$. The syndrome-based messaging allows a 2-bit communication. To encode secret messages, Alice applies her errors according to the convention given in Table \ref{table:1}.
\begin{table}
\caption{Convention according to which Alice applies her errors.}
\begin{center}
    \begin{tabular}{|c|c|c|}
\hline Pauli operator & Error applied & Degenerate counterpart\\
\hline
   $I$  & $I$ & $I$ \\
   \hline
   $X$  & $Z_1, Z_2$ & $Z_3$\\
   \hline
   $Y$ & $Z_4, Z_5$ & $Z_6$\\
   \hline
   $Z$ & $Z_7, Z_8$ & $Z_9$\\
   \hline
\end{tabular}
\end{center}
\label{table:1}
\end{table}
Thus $\mathfrak{E}_A = \{I, Z_1, Z_2, Z_4, Z_5, Z_6, Z_7\}$. Suppose Bob's decoding reveals an error $Y$. He infers that the transmitted state must be $Y\ket{\varphi}$, and corrects it by applying the operator. In the event of Eve's challenge, Bob randomly applies error from the set $\mathfrak{E}_B=\{I, Z_3, Z_6, Z_9\}$.

To transmit a secret state $\ket{\varphi}$, Alice applies a random Pauli error $P_i$, and $P_i\ket{\varphi}$ will be the cover message. In the challenge mode, to erase or randomize the secret key, Bob randomly applies one of $I, Z_3, Z_6, Z_9$ on his qubits before surrendering his three qubits to Eve. If Bob applies $Z_6$, which acts on the block as $Z_4$, then the errors cancel by virtue of degeneracy, noting that $Z_6Z_4$ lies in the stabilizer $\mathcal{S}$. On the other hand, he could also apply $Z_3$ or $Z_9$, which lie in a different block than $Z_4$. Suppose without loss of generality, it is the latter:
 \begin{align}
 Z_9 Z_4\ket{0_L} = (Z_1)^2 Z_9 Z_4\ket{0_L} =Z_1 X_L \ket{0_L} =Z_1\ket{1_L}.
 \label{eq:error-state}
 \end{align}
Thus, Eve interprets the state as a $Z_1$ error on the code word $\ket{1_L}$. Moreover, without knowing the decryption key, Eve essentially gets the randomized message transmitted by Bob.

Last but not least, it is important to note that the reduced state with Alice is
$$ (\ketbra{00}{}{00} + \ketbra{11}{}{11})^{\otimes 3},$$
irrespective of the encoded state $\ket{\varphi}$ or the injected error syndromes.

\section{Steganography using the phase bit of ebit}
\label{sec:scheme-3}
In the stego protocol of Ref. \cite{mihara2015quantum}, the parity bit of a preshared EPR pair is used for transmitting the secret bit, reminiscent of partial dense coding.  The encoding employs parts of a quantum code rather than the full code. Yet parity errors can be corrected by applying the parity check matrix of the underlying classical code, to each ket in the superposition. The receiver corrects the phase flip error later, after making the parity-error corrected state separable. If we wish to use the phase bit instead of the parity bit for the purpose, directly adapting Mihara's idea is not possible, because the underlying classical code cannot correct phase errors, and converting the phase errors to bit-flip errors by applying Hadamards to all qubits will not work because the state with Bob is not separable. Thus, we need to make suitable modifications, as described below.

\subsection{Protocol 3}
A linear $[[n,k,d]]$ QECC can be defined by $m=n-k$  stabilizer generators. Let $\ket{w_L}$ $(0 \le w \le 2^k-1)$ denote a codeword of this QECC, and $q \equiv \lfloor\frac{d-1}{2} \rfloor +1$. Let $\mathbb{G}$ be the group consisting of all computational basis states in the support of $\ket{0_L}$ and let $L_0$ $(\subseteq \mathbb{G})$ be the subgroup defined as follows:
\begin{equation}
    L_0 = \{v \in \mathbb{G} ~|~ \Vec{q}\cdot v=0\},
\end{equation}
where $\Vec{q}$ is a $n$-bit vector of Hamming weight $|\Vec{q}|=q$. Applying the Lagrange theorem, we find that the integer $l \equiv \frac{\abs {\mathbb{G}}}{\abs {L_0}} \in \{1,2\}$. We choose $\Vec{q}$ such that $l=2$. Further define coset $L_1 \equiv \mathbb{G} - L_0$. Note that $\forall_{v \in L_1} \Vec{q} \cdot v=1$.

We express $\ket{0_L}$ as the superposition
\begin{equation}
    \ket{0_L}=\frac{1}{\sqrt{2}} \bigl(\ket{L_0} + \ket{L_1} \bigr),
    \label{eq:split}
\end{equation}
where
\begin{align}
    \ket{L_j} &= \sqrt{2}\big( \sum_{v \in L_j}\ket{v}\braket{v|0_L}\big),~~(j\in\{0,1\}).
    \label{eq:splitting}
\end{align}
Analogous to Eq. (\ref{eq:split}), we can define such superpositions for all codewords:
\begin{equation}
    \ket{w_L}=\frac{1}{\sqrt2} \bigl(\ket{L_0^{(w)}} + \ket{L_1^{(w)}} \bigr),
    \label{eq:splitw}
\end{equation}
since in each case, $L_0^{(w)} \cup L_1^{(w)}$ is a coset of $\mathbb{G}$. Note that in our notation $L_j^{(0)}=L_j$ for $j \in \{0,1\}$.

Following are protocol steps,
\begin{enumerate}
    \item Alice and Bob share an entangled state
\begin{align}
 \ket{\Phi^+} =  \frac{1}{\sqrt{2}}\bigl(\ket{0,0}+ \ket{1,1}\bigr)_{AB}.
\end{align}
For encoding secret bit $b\in \{0,1\}$, she applies the local unitary $Z^b_1$ to her qubit to produce the state
\begin{align}
    Z^b_1 \ket{\Phi^+} = \frac{1}{\sqrt{2}}\bigl(\ket{0,0}+ (-1)^b\ket{1,1}\bigr)_{AB}.
    \label{eq:Phib}
\end{align}
\item She encodes her entangled qubit to create the following state:
\begin{align}
  \ket{\Upsilon(w,b)} \equiv \frac{1}{\sqrt{2}} \left(\ket{L^{(w)}_0,0} + (-1)^b\ket{L^{(w)}_{1},1}\right)_{AB}.
  \label{eq:stego-message}
\end{align}
\item She dispatches her qubits to Bob. During transit, her $n$ qubits may pick up an arbitrary error $\mathfrak{E}$ of weight up to $\lfloor \frac{d-1}{2}\rfloor$. 
\item Bob receives the erroneous state
\begin{align}
    \mathfrak{E}\ket{\Upsilon(w,b)} = \mathfrak{E}\frac{1}{\sqrt{2}}\bigl(\ket{L^{(w)}_0,0}+(-1)^b \ket{L^{(w)}_1,1}\bigr)_{AB}.
    \label{eq:Esteog-message}
\end{align}
Bob's decoding procedure is described below.
\end{enumerate}
In general, Bob will not be able to measure the error syndrome, because the code fragments $\ket{L^{(w)}_j}$ ($j \in \{0,1\}$) are entangled with Bob's qubit. Error correction will thus require an indirect method.

We define the Hamming support $h_q(\mathcal{P})$ of an $n$-qubit Pauli operator $\mathcal{P}$ as the set of Pauli operators that appear in $\mathcal{P}$ at the coordinates where $\hat{q}$ has 1. For example, given $\hat{q} = (1,0,0,1)$, we have $\hat{q}[IXYX] = \{I_1,X_4\}$ and $\hat{q}[ZXIZ] = \{Z_1,Z_4\}$.
Now, the set of stabilizer generators is split into two classes.
\begin{description}
    \item[Non-flipping] A stabilizer generator $S_{\rm NF}$ such that $h_q(S_{\rm NF})$ contains an even number of operators $X$ and $Y$. It is easy to show that they have the symmetry property
\begin{equation}
      S_{\rm NF} \ket{L^{(w)}_j}=\ket{L^{(w)}_j}~~(j \in \{0,1\}).
      \label{eq:nf-stab}
    \end{equation}
    In other words, the stabilizers $S_{\rm NF}$ are equivalent to an identity operation in the subspace $\mathfrak{S}^{(w)}$ spanned by the vectors $\ket{L^{(w)}_0}$ and $\ket{L^{(w)}_1}$.
    \item[Flipping] A stabilizer generator $S_{\rm F}$ is such that $h_q(S_{\rm F})$ contains an odd number of operators $X$ or $Y$. It is easy to show that they have the flipping property
\begin{align}
     S_{\rm F} \ket{L^{(w)}_j}=\ket{L^{(w)}_{\overline{j}}}~~(j \in \{0,1\}),
      \label{eq:f-stab}
    \end{align}
where the overline represents the complementation operation $\overline{0}=1$ and $\overline{1}=0$. In other words, the stabilizers $S_{\rm F}$ are equivalent to a NOT operation in the subspace $\mathfrak{S}^{(w)}$. 
\end{description}

The syndrome corresponding to a non flipping stabilizer generator can be obtained in the standard way, by ignoring the entanglement. Given stabilizer $S_{\rm NF}$ with corresponding syndrome $s_{\rm NF}$, we find
\begin{align}
    (S_{\rm NF} \otimes \mathbb{I})&(\mathfrak{E} \otimes \mathbb{I}) \ket{\Upsilon(w,b)} \nonumber \\ &= (S_{\rm NF}\mathfrak{E} \otimes \mathbb{I})\frac{1}{\sqrt{2}}\bigl(\ket{L^{(w)}_0,0}+(-1)^b \ket{L^{(w)}_1,1}\bigr)_{AB} \nonumber\\
    &= (-1)^{s_{\rm NF}}(\mathfrak{E} \otimes \mathbb{I})\frac{1}{\sqrt{2}}\big(\ket{L^{(w)}_0,0}+(-1)^b \ket{L^{(w)}_1,1}\big)_{AB}~~,
    \label{eq:nf-correction}
\end{align}
where $s_{\rm NF} (\in \{0,1\})$. 

However, the state in Eq. (\ref{eq:stego-message}) is not an eigenstate of any $(S_{\rm F} \otimes \mathbb{I})$. Thus, error correction must proceed in another way. To this end, we observe that the stego state is an eigenstate of the product of any two distinct flipping stabilizer generators, for
\begin{align*}
 (S_{\rm F}^{\prime}S_{\rm F} \mathfrak{E} &\otimes \mathbb{I}) \ket{\Upsilon(w,b)}_{AB} \nonumber \\
 &=(S_{\rm F}^{\prime}S_{\rm F} \mathfrak{E} \otimes \mathbb{I})\frac{1}{\sqrt{2}}\bigl(\ket{L^{(w)}_0,0}+(-1)^b \ket{L^{(w)}_1,1}\bigr)_{AB} \nonumber \\
    &= (-1)^{s_{\rm F}^{\prime} + s_{\rm F}} (\mathfrak{E} \otimes \mathbb{I})\frac{1}{\sqrt{2}}\big(\ket{L^{(w)}_0,0}+(-1)^b \ket{L^{(w)}_1,1}\big)_{AB}~~ \nonumber \\
\end{align*}
where $s_{\rm F},s^{'}_{\rm F} \in \{0,1\}$.
Bob extracts the syndrome value $s_{\rm F}^{\prime} + s_{\rm F}$. Proceeding thus pairwise, he can obtain ${\varphi \choose 2}$ sums, where $\varphi$ is the number of flipping stabilizers. Provided this is not smaller than $\varphi$, i.e., $${\varphi \choose 2} \ge \varphi,$$ we have enough sums to simultaneously solve for the $\varphi$ eigenvalues of the flipping stabilizer generators.

From the error syndromes he can identify $\mathfrak{E}$ and correct it to obtain the original state Eq. (\ref{eq:stego-message}). To this end, we are required to identify the \textit{sublogical} operations, i.e., encoded operations for the subspace spanned by $\{\ket{L_0}, \ket{L_1}\}$. Denote by $q_i$ $(1 \le i \le n)$ the bits constituting vector $\Vec{q}$. Then the encoded $Z$ operator in the subspace spanned by $\ket{L_0}$ and $\ket{L_1}$, denoted $\overline{Z}$, is given by $\bigotimes_{i=1}^n Z^{q_i}$, with the identity $Z^0=\mathbb{I}$. For example, given $\Vec{q}=010011$, we have $\overline{Z}=Z_2Z_5Z_6$. This ensures that $\overline{Z} \ket{L_0}= \ket{L_0}$, $\overline{Z} \ket{L_1}= -\ket{L_1}$. Recollecting that the flipping and nonflipping stabilizers behave, respectively, like the encoded identity $\overline{I}$ and $\overline{X}$ in the subspace $\mathfrak{S}^{(w)}$, we find that $[\overline{Z},S_{\rm NF}]=0$ and $\{\overline{Z},S_{\rm F}\}=0$.

Now Bob applies controlled-$S_{\rm F}$  to the stego state in Eq. (\ref{eq:stego-message}), with the transmitted qubits as target and his entangled qubit as control:
\begin{align}
    \ket{\Upsilon(w,b)}_{AB} \xrightarrow[]{C-S_{\rm F}} \ket{L^{(w)}_0} \big(\ket{0}+(-1)^b\ket{1}\big).
    \label{eq:Cont-F}
\end{align}
Measuring his qubit in the $X$ basis, he extracts the secret message $b$. Finally, application of the \textit{sublogical} Hadamard $\overline{H} \equiv \frac{1}{\sqrt{2}}(S_{\rm F} + \overline{Z})$ to the transmitted qubits restores the cover message 
\begin{equation}
\ket{L^{(w)}_0} \xrightarrow[]{\overline{H}} \frac{1}{\sqrt{2}}(\ket{L_0^{(w)}} + \ket{L_1^{(w)}})  \equiv \ket{w_L},
\label{eq:unsplit}
\end{equation}
by Eq. (\ref{eq:splitw}).
We remark that by virtue of linearity, the secret can be a quantum state $\alpha\ket{b=0} + \beta\ket{b=1}$, $|\alpha|^2 + |\beta|^2=1$, in which case the initial entangled state will be $\alpha\ket{\Phi^+} + \beta\ket{\Phi^-}$ in Eq. (\ref{eq:Phib}). Similarly, the cover message can also be a superposition $\sum_w c_w\ket{w}$ $(\sum_w |c_w|^2=1)$. 

\subsection{Secrecy, security, and capacity bounds}
As with our preceding stego protocols, secret messages are stored in the nonlocal correlations rather than error syndromes. Thus a presumption of a degree of Eve's ignorance of the channel or fine-tuning the secret messaging is not important.
From Eq. (\ref{eq:stego-message}), we note that the state of the particle in transit is given by
\begin{align}
    \text{Tr}_B\big(\mathfrak{E} \ketbra{\Upsilon(w,b)}{AB}{\Upsilon(w,b)} \mathfrak{E}^{\dagger}\big) &= \frac{1}{2}\big(\mathfrak{E}\ketbra{L^{(w)}_0}{A}{L^{(w)}_0}\mathfrak{E}^{\dagger} \nonumber \\ &+ \mathfrak{E}\ketbra{L^{(w)}_1}{A}{L^{(w)}_1}\mathfrak{E}^{\dagger}\big),
\end{align}
irrespective of the secret bit $b$ and depending possibly only on the cover message $w$ and noise $\mathfrak{E}$, in keeping with the secrecy condition Eq. (\ref{eq:equal}). Thus Eve the eavesdropper detects no suspicious patterns in the state transmission. Note that this is analogous to Eq. (\ref{eq:secrecy1}) in the case of the dense-coding protocol 1. 

To be specific, consider the following entangling attack by Eve on the above protocol. Eve randomly inspects some transmissions from Alice, by intercepting them. Once given such a transmission, she performs a suitable unitary operation $U_{AE}$ to entangle it with her probe system $E$ prepared in state $\ket{\varepsilon}$ in an effort to implement a phase extraction mechanism. For the states $\ket{\Upsilon(w,b)}$, suppose
\begin{align}
U_{AE}\ket{\Upsilon(w,b)}_{AB}|\varepsilon\rangle_E &= \frac{1}{\sqrt{2}}(|L_0^{(w)},0,\varepsilon_{0}\rangle + (-1)^b|L_1^{(w)},1,\varepsilon_{1}\rangle)_{ABE}, \nonumber \\
&\equiv \ket{\Upsilon^{\bowtie}(w,b)}_{ABE}
\end{align}
where $\ket{\varepsilon_0}$ and $\ket{\varepsilon_1}$ are in general not orthogonal. Eve hopes to discern the phase by measuring her probe in a suitable basis. She expects to encounter an information-disturbance tradeoff whereby to gain greater information about phase, a large disturbance ($|\braket{\varepsilon_0|\varepsilon_1}|\approx0$) would be produced and, conversely, a small disturbance ($|\braket{\varepsilon_0|\varepsilon_1}|\approx1$) would result in smaller information gain.

The key point is that the probe state is given by $ {\rm Tr}_{AB}\big(\ketbra{\Upsilon^{\bowtie}(w,b)} {ABE} {\Upsilon^{\bowtie}(w,b)}\big)= \rho_E^{(b)} = \frac{1}{2}(\ketbra{\varepsilon_0}{E}{\varepsilon_0} + \ketbra{\varepsilon_1}{E}{\varepsilon_1}$, where the phase $(-1)^b$ is seen to drop out entirely from $\rho_E^{(b)}$. Therefore, having only local access to $A$, Eve gains zero information about the phase bit: $I(b : E) = 0$. As such, the question of an information-disturbance tradeoff does not arise.

Regarding security, we note from Eqs. (\ref{eq:Cont-F}) and (\ref{eq:unsplit}) that Bob effectively performs the following action on the error corrected state:
\begin{align}
    \ket{\Upsilon(w,b)}_{AB} \xrightarrow{\overline{H}\otimes H\circ C-S_{\rm F}} \ket{w_L}\ket{b},
\end{align}
where the information in the first register $(w)$ in the right-hand side is handed over to Eve as the cover in the event of being challenged by her.

The present protocol is conceptually similar to protocol 1, except that the encoding and decoding procedures are different. Both these protocols correspond to a dense-coding-like strategy where half of the transmitted classical information serves as the secret bit. Thus for the depolarizing channel from Eqs. (\ref{eq:capdepol0}) and (\ref{eq:floordepol0}) the lower and upper bounds are given by
\begin{equation}
1 - 2p\log(3) - H\left(2p\right) \le
    R_s^C \le 1 - \frac{p}{2}\log(3) - \frac{1}{2}H\left(p\right).
\label{eq:capsdepol1}
\end{equation}
Analogously, for the dephasing channel, we find from Eqs. (\ref{eq:capdephas}) and (\ref{eq:stegoGV})
\begin{equation}
1 - H_2(2p) \le    R_s^C \le 1 - \frac{1}{2}H(p).
    \label{eq:3dephas}
\end{equation}
Purely by rate consideration, the present protocol is as robust as protocol 1. However, the practical implementations of the multiqubit operators are different from those in standard quantum error correction, and would require specific circuits. The main interest of the protocol lies in the theoretical idea of storing the secret as the phase bit (rather than the parity bit, as in Ref. \cite{mihara2015quantum}) of split codewords. 
\color{black}

\subsection{Example}
Below we present an example of this protocol using the five-qubit code. Consider the $[[n=5,k=1,d=3]]$ QECC code for steganographic communication defined by the stabilizer generators
$S_1= XZZXI,  S_2= XIXZZ, S_3=IXZZX,   S_4=ZXIXZ$. For the cover message $w \in \{0,1\}$, we use the encoded cover message state
$\ket{w_L}= \prod_{i=1}^{4}(I^{\otimes5} + S_i)\ket{w}^{\otimes5}$ \cite{devitt2013quantum}. For the cover message $w=0$, the logical state is
\begin{align}
      \ket{0_L}  & =\frac{1}{4} \Bigl(\ket{00000}+\ket{10010}+\ket{01001}
      +\ket{10100} \nonumber \\ &+\ket{01010}- \ket{11011}-\ket{00110}-\ket{11000}\nonumber\\
      & -\ket{11101}-\ket{00011}-\ket{11110}-\ket{01111}\nonumber\\ &-\ket{10001}-\ket{01100}
      -\ket{10111}+\ket{00101}\Bigr).
      \label{eq:five-qubit}
 \end{align}
We choose $\Vec{q}=(0,0,0,1,1)$. In the manner of Eqs. (\ref{eq:split}) and (\ref{eq:splitting}), we have 
\begin{align}
  \ket{0_L}
  =\frac{1}{\sqrt{2}} \Bigl(\ket{L_0} + \ket{L_1}\Bigr),
 \end{align}
where
 \begin{align}
   \ket{L_0}= &\frac{1}{2\sqrt{2}}\Bigl(\ket{00000}+\ket{10100}- \ket{11011}-\ket{11000}\nonumber\\ &-\ket{00011}-\ket{01111}-\ket{01100} -\ket{10111}\Bigr),  \nonumber \\
 \ket{L_1}
  =&\frac{1}{2\sqrt{2}} \Bigl(\ket{10010}+\ket{01001}+\ket{01010}-\ket{00110}\nonumber\\&-\ket{11101}-\ket{11110}-\ket{10001}+\ket{00101}\Bigr).   
\end{align}
Beforehand, sender Alice and receiver Bob pre-share the Bell state $\ket{\Phi^+}$. Alice encodes the secret message $b$ as in Eq. (\ref{eq:Phib}). She then encodes her particle in the manner of Eq. (\ref{eq:stego-message}), $\ket{\Upsilon(0,b)}_{AB} \equiv \frac{1}{\sqrt{2}} \left(\ket{L_0}\ket{0} + (-1)^b\ket{L_{1}} \ket{1}\right)_{AB}$.  She transmits her five qubits to Bob through the noisy channel and Bob receives the erroneous state as given in Eq. (\ref{eq:Esteog-message}) $\mathfrak{E}\ket{\Upsilon(w,b)}$, with $\mathfrak{E}$ being an arbitrary single-qubit error. We assume that his entangled particle is error free, an assumption justified because his qubit has not been transmitted across the channel.  

Here $S_1,S_3$, and $S_4$ are the flipping stabilizers, while $S_2$ is the only nonflipping stabilizer. Upon receiving Alice's qubits, Bob implements the following protocol.

(1) For the nonflipping stabilizer, Bob can extract the syndrome by direct measurement of the stabilizer on the first five qubits: 
    \begin{align}
        S_2\mathfrak{E}\ket{\Upsilon(0,b)}_{AB}&=S_2 \mathfrak{E}(\ket{L_0}\ket{0} + (-1)^b\ket{L_1}\ket{1})\nonumber\\ &= (-1)^{s_2} \mathfrak{E} S_2(\ket{L_0}\ket{0} + (-1)^b\ket{L_1}\ket{1}) \nonumber\\
        &= (-1)^{s_2} \mathfrak{E} \ket{\Upsilon(0,b)}_{AB}.
    \end{align}
    
(2) For the flipping stabilizers, Bob extracts the pairwise sums of syndromes by joint measurement of pairs of stabilizers. For example,
    \begin{align}
        S_4 S_3 \mathfrak{E}\ket{\Upsilon(0,b)}_{AB}&= S_4 S_3 \mathfrak{E}(\ket{L_0}\ket{0} + (-1)^b\ket{L_1}\ket{1})_{AB} \nonumber \\ &= (-1)^{s_3} S_4 \mathfrak{E} S_3(\ket{L_0}\ket{0} + (-1)^b\ket{L_1}\ket{1})_{AB} \nonumber\\
        &= (-1)^{s_3} S_4 \mathfrak{E} (\ket{L_1}\ket{0} + (-1)^b\ket{L_0}\ket{1})_{AB}, \nonumber\\
        &= (-1)^{s_3+s_4} \mathfrak{E} S_4 (\ket{L_1}\ket{0} + (-1)^b\ket{L_0}\ket{1})_{AB} \nonumber\\
        &= (-1)^{s_3+s_4} \mathfrak{E} \ket{\Upsilon(0,b)}_{AB}.
    \end{align}
By measuring $S_3 S_4$, Bob obtains the sum of syndromes, or $s_3 + s_4$.

(3) By simultaneously solving the syndrome sums $s_3+s_4, s_1+s_3, s_1+s_4$, Bob extracts the individual $s_j$'s, and thereby determines and corrects error $\mathfrak{E}$. 

(4) He applies a controlled $S_{\rm F}$ as in Eq. (\ref{eq:Cont-F}):
\begin{align}
  \ket{\Upsilon(0,b)}_{AB} 
  &\xrightarrow{\overline{\text{CNOT}}} \ket{L_0}\frac{1}{\sqrt{2}}(\ket{0} + (-1)^b\ket{1}),
  \label{eq:nihara}
\end{align} 
and determines $b$ by measuring his qubit in the $X$ basis.

(5) In order to make the cover message available, the \textit{sublogical} Hadamard $\overline{\mathcal{H}}=\frac{1}{\sqrt{2}} \Bigl(S_1 + Z_4Z_5\Bigr)$ is applied to the code qubits, which effects the transformation
$$ \ket{L_0} \xrightarrow{\overline{\mathcal{H}}} \frac{1}{\sqrt{2}} \left(\ket{L_0} + \ket{L_{1}}\right) \equiv \ket{0_L}.$$
Note that $Z_4Z_5$ is the sublogical $Z$ operator, and any other flipping stabilizer could be used in place of $S_1$. 

\section{Conclusions and discussions \label{sec:conclusion}}
In this work, we presented protocols for quantum steganography that employ catalytic, entanglement-assisted quantum codes, or quantum nonlocality,  aiming for greater resource efficiency and newer encoding strategies. The protocols are designed to transmit secret messages by hiding them either within typical error patterns dictated by QECCs or in quantum nonlocal correlations, leveraging properties unique to quantum channels such as superposition, no-cloning, and entanglement for guarantee of secrecy and security.

Our first protocol introduces the concept of catalytic quantum codes in steganography, enabling the recycling of preshared entanglement resources rather than requiring fresh ebits for each round of steganographic communication. This significantly reduces the quantum overhead and makes repeated communication more realistic. By applying degenerate entanglement-assisted codes to steganography, the second protocol enables both sender and receiver to actively contribute to enhanced resilience against eavesdropping and for flexible embedding of secret data. Building on Ref. \cite{mihara2015quantum}, our final protocol utilizes quantum nonlocal correlations like the first protocol, but employing a nonstandard encoding, leading to modified ways to correct errors and extract hidden messages securely. In each of these cases, we show how transmissions remain indistinguishable from normal quantum traffic, thus minimizing the risk of steganalysis by adversaries. This multipronged approach provides practical strategies for embedding secrets in quantum datastreams while maintaining error correction, resource replenishment, and concealment from eavesdroppers.

\begin{figure}[htp]
    \centering
    \includegraphics[width=8cm]{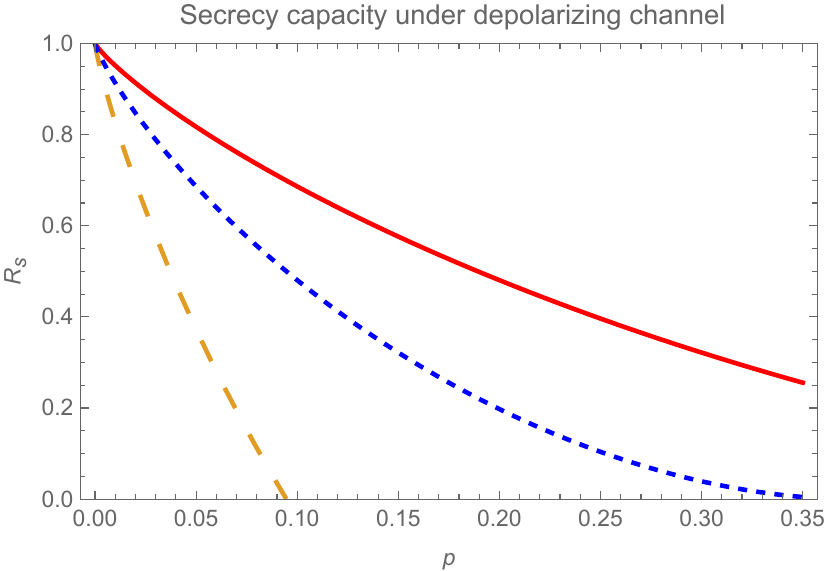}
    \caption{(Color online) Upper bound (solid line) and lower bounds (small- and large-dashed lines) on the classical secrecy capacity for Protocol 1 (or 3) and the quantum secrecy capacity for Protocol 2 under the depolarizing channel.  The upper bounds coincide, depending only on the noise parameters, whereas Protocol 2 gives a better lower bound.}
      \label{fig:depol}
     \end{figure}

\begin{figure}[htp]
    \centering
    \includegraphics[width=8cm]{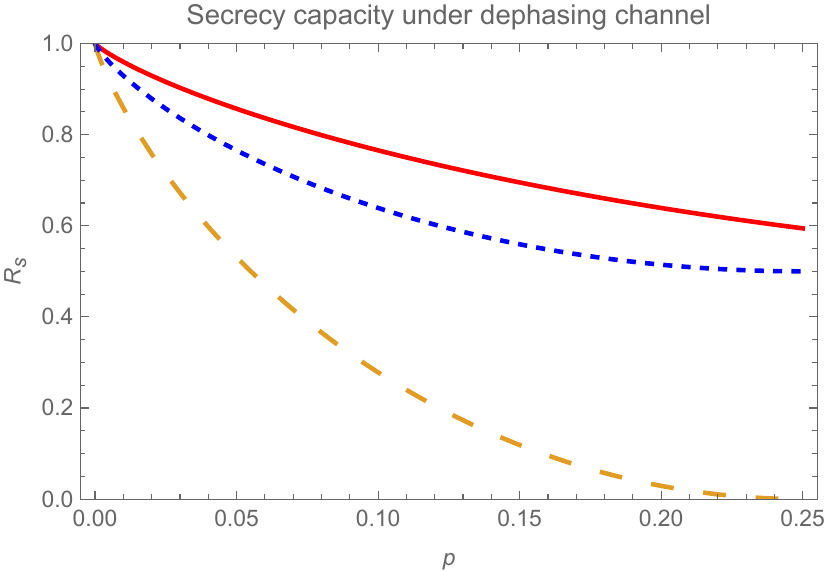}
    \caption{(Color online) Upper bound (solid line) and lower bounds (small- and large-dashed lines) on the classical secrecy capacity for protocol 1 (or 3) and the quantum secrecy capacity 2 under the dephasing channel.  The upper bounds coincide, whereas Protocol \#2 gives a better lower bound, as in Fig. \ref{fig:depol}.}
      \label{fig:dephas}
     \end{figure}

In the context of practical deployment of quantum steganography, a comparison of our protocols' performances is useful. For the considered noise models, in terms of the secrecy capacity and robustness, protocols 1 and 3 are similar in performance, while that of Protocol 2 deviates from them, as depicted in Figs. \ref{fig:depol} and \ref{fig:dephas}, respectively, for the depolarizing and dephasing channels. For the former channels, the upper and lower bounds on classical secrecy capacities are given by Eqs. (\ref{eq:capdepol0}) and (\ref{eq:floordepol0}) for Protocols 1 or 3, while the corresponding quantum upper and lower bounds are given by Eqs. (\ref{eq:capdepol}) and (\ref{eq:lowercapacity1}) for protocol 2.

Figure \ref{fig:depol} shows that for a given noise level $p$, while the upper bound of classical secrecy capacity of protocols 1 or 3 coincides with that for the quantum secrecy capacity of protocol 1, yet protocol 2 gives a larger lower bound on the secrecy capacity. For the dephasing channel, the upper and lower bounds on secrecy capacities are given by Eqs. (\ref{eq:capdephas}) and (\ref{eq:stegoGV}) for protocols 1 or 3, while the corresponding upper and lower bounds are given by Eqs. (\ref{eq:capdeph1}) and (\ref{eq:lowercapacity3}) for protocol 2. 
     
Figure \ref{fig:dephas} shows that while the upper bound of secrecy capacity coincides for all three protocols, Protocol \#2 gives a larger lower bound on secrecy capacity, as in the case of the depolarizing channel. In terms of the capacity bounds and the ability to send quantum rather than classical covert information, protocol 2 offers a better performance, which can be attributed to the use of EA encoding. However, the structure of codes used here requires suitable steganographic and secret-sharing properties, as discussed. 

Our protocols can find practical applications in a number of current scenarios, among them the following. (a) \textit{Military covert channels:} Protocol 2 can be used to embed classified commands in quantum network traffic  \cite{agrawal2024perspective} appearing as error correction overhead. (b) \textit{Healthcare records}  \cite{abd2020robust}: Protocol 1 can be used to conceal patient data in quantum states sent between hospitals sharing entanglement; the correlations are invisible to unauthorized access. (c) \textit{Financial transactions:} Protocol 2 can be embedded in quantum cloud storage or blockchain oracles to hide transaction details. (d) \textit{Smart grid security:} Steganographic commands in power grid quantum sensors mask against cyber threats, for which Protocol 1 or 2 can be used.

It is worth considering the challenges and limitations associated with a practical realization of these protocols. Quantum steganography minimally requires the ability to implement quantum communication, which entails preparation, transmission, and measurement of quantum states. But going beyond this, the protocols presented here require presharing and holding a sufficient number of entangled pairs, which will require quantum memory and the capacity to implement entanglement distribution. It is worth noting that we have assumed the preshared entanglement to be noiseless, which will be difficult to guarantee in practice, and requires one to be relaxed in any real-life practical setting. 

The above considerations suggest various possible future directions, mainly in the direction of refining these protocols and for higher communication rates. An interesting challenge would be to explore the integration of the above protocols, or indeed any stego protocol, into quantum secure direct communication formats \cite{banerjee2012maximally, sheng2022one}. On another front, it would seem that we cannot replace the role of QECCs with decoherence-free subspaces (DFS) \cite{lidar1998decoherence}, which remain invariant under the action of certain error operators. However, in the manner of the role of degeneracy in the second protocol above, DFS may bring unexpected benefits to steganography. As such, this issue needs to be explored, and corroborated or refuted. From the perspective of practical implementation, the use of continuous-variable codes such as the Gottesman-Kitaev-Preskill code \cite{gottesman2001encoding, grimsmo2021quantum} instead of QECCs in the context of stego protocols is worth studying, though the scope for embedding secret messages as errors seems limited. A natural future direction of investigation would be to expand from stego protocols focused on point-to-point communication to a quantum network. Here the use of concatenations of quantum codes \cite{dash2024concatenating} and
surface codes \cite{kitaev2003fault, fowler2012surface} can prove advantageous. \color{black}

\acknowledgements
S.D. acknowledges the financial assistance of a UGC NET scholarship and also Udupi Sri Admar Mutt Education Foundation. N.R.D. acknowledges financial support from the Department of Science and Technology, Ministry of Science and Technology, India, through the INSPIRE fellowship. R.S. acknowledges the partial financial support of the Indian Science and Engineering Research Board (Grant No. CRG/2022/008345). 
\bibliography{reference}

\begin{thebibliography}{43}%
\makeatletter
\providecommand \@ifxundefined [1]{%
 \@ifx{#1\undefined}
}%
\providecommand \@ifnum [1]{%
 \ifnum #1\expandafter \@firstoftwo
 \else \expandafter \@secondoftwo
 \fi
}%
\providecommand \@ifx [1]{%
 \ifx #1\expandafter \@firstoftwo
 \else \expandafter \@secondoftwo
 \fi
}%
\providecommand \natexlab [1]{#1}%
\providecommand \enquote  [1]{``#1''}%
\providecommand \bibnamefont  [1]{#1}%
\providecommand \bibfnamefont [1]{#1}%
\providecommand \citenamefont [1]{#1}%
\providecommand \href@noop [0]{\@secondoftwo}%
\providecommand \href [0]{\begingroup \@sanitize@url \@href}%
\providecommand \@href[1]{\@@startlink{#1}\@@href}%
\providecommand \@@href[1]{\endgroup#1\@@endlink}%
\providecommand \@sanitize@url [0]{\catcode `\\12\catcode `\$12\catcode `\&12\catcode `\#12\catcode `\^12\catcode `\_12\catcode `\%12\relax}%
\providecommand \@@startlink[1]{}%
\providecommand \@@endlink[0]{}%
\providecommand \url  [0]{\begingroup\@sanitize@url \@url }%
\providecommand \@url [1]{\endgroup\@href {#1}{\urlprefix }}%
\providecommand \urlprefix  [0]{URL }%
\providecommand \Eprint [0]{\href }%
\providecommand \doibase [0]{https://doi.org/}%
\providecommand \selectlanguage [0]{\@gobble}%
\providecommand \bibinfo  [0]{\@secondoftwo}%
\providecommand \bibfield  [0]{\@secondoftwo}%
\providecommand \translation [1]{[#1]}%
\providecommand \BibitemOpen [0]{}%
\providecommand \bibitemStop [0]{}%
\providecommand \bibitemNoStop [0]{.\EOS\space}%
\providecommand \EOS [0]{\spacefactor3000\relax}%
\providecommand \BibitemShut  [1]{\csname bibitem#1\endcsname}%
\let\auto@bib@innerbib\@empty
\bibitem [{\citenamefont {Simmons}(1984)}]{simmons1984prisoners}%
  \BibitemOpen
  \bibfield  {author} {\bibinfo {author} {\bibfnamefont {G.~J.}\ \bibnamefont {Simmons}},\ }\bibfield  {title} {\bibinfo {title} {The prisoners’ problem and the subliminal channel},\ }in\ \href@noop {} {\emph {\bibinfo {booktitle} {Advances in Cryptology: Proceedings of Crypto 83}}}\ (\bibinfo {organization} {Springer},\ \bibinfo {year} {1984})\ pp.\ \bibinfo {pages} {51--67}\BibitemShut {NoStop}%
\bibitem [{\citenamefont {Avritzer}\ and\ \citenamefont {Brun}(2024)}]{avritzer2024quantum}%
  \BibitemOpen
  \bibfield  {author} {\bibinfo {author} {\bibfnamefont {B.}~\bibnamefont {Avritzer}}\ and\ \bibinfo {author} {\bibfnamefont {T.~A.}\ \bibnamefont {Brun}},\ }\bibfield  {title} {\bibinfo {title} {Quantum steganography via coherent-and fock-state encoding in an optical medium},\ }\href@noop {} {\bibfield  {journal} {\bibinfo  {journal} {Physical Review A}\ }\textbf {\bibinfo {volume} {109}},\ \bibinfo {pages} {032401} (\bibinfo {year} {2024})}\BibitemShut {NoStop}%
\bibitem [{\citenamefont {Dutta}\ \emph {et~al.}(2020)\citenamefont {Dutta}, \citenamefont {Das}, \citenamefont {Nandi},\ and\ \citenamefont {Prasanna}}]{dutta2020overview}%
  \BibitemOpen
  \bibfield  {author} {\bibinfo {author} {\bibfnamefont {H.}~\bibnamefont {Dutta}}, \bibinfo {author} {\bibfnamefont {R.~K.}\ \bibnamefont {Das}}, \bibinfo {author} {\bibfnamefont {S.}~\bibnamefont {Nandi}},\ and\ \bibinfo {author} {\bibfnamefont {S.~M.}\ \bibnamefont {Prasanna}},\ }\bibfield  {title} {\bibinfo {title} {An overview of digital audio steganography},\ }\href@noop {} {\bibfield  {journal} {\bibinfo  {journal} {IETE Technical Review}\ }\textbf {\bibinfo {volume} {37}},\ \bibinfo {pages} {632} (\bibinfo {year} {2020})}\BibitemShut {NoStop}%
\bibitem [{\citenamefont {Hu}\ \emph {et~al.}(2020)\citenamefont {Hu}, \citenamefont {Zhou}, \citenamefont {Liu}, \citenamefont {Luo},\ and\ \citenamefont {Luo}}]{hu2020quantum}%
  \BibitemOpen
  \bibfield  {author} {\bibinfo {author} {\bibfnamefont {W.-W.}\ \bibnamefont {Hu}}, \bibinfo {author} {\bibfnamefont {R.-G.}\ \bibnamefont {Zhou}}, \bibinfo {author} {\bibfnamefont {X.-A.}\ \bibnamefont {Liu}}, \bibinfo {author} {\bibfnamefont {J.}~\bibnamefont {Luo}},\ and\ \bibinfo {author} {\bibfnamefont {G.-F.}\ \bibnamefont {Luo}},\ }\bibfield  {title} {\bibinfo {title} {Quantum image steganography algorithm based on modified exploiting modification direction embedding: W. hu et al.},\ }\href@noop {} {\bibfield  {journal} {\bibinfo  {journal} {Quantum Information Processing}\ }\textbf {\bibinfo {volume} {19}},\ \bibinfo {pages} {137} (\bibinfo {year} {2020})}\BibitemShut {NoStop}%
\bibitem [{\citenamefont {Liu}\ \emph {et~al.}(2021)\citenamefont {Liu}, \citenamefont {Ni}, \citenamefont {Zhang},\ and\ \citenamefont {Huang}}]{liu2021novel}%
  \BibitemOpen
  \bibfield  {author} {\bibinfo {author} {\bibfnamefont {Y.}~\bibnamefont {Liu}}, \bibinfo {author} {\bibfnamefont {J.}~\bibnamefont {Ni}}, \bibinfo {author} {\bibfnamefont {W.}~\bibnamefont {Zhang}},\ and\ \bibinfo {author} {\bibfnamefont {J.}~\bibnamefont {Huang}},\ }\bibfield  {title} {\bibinfo {title} {A novel video steganographic scheme incorporating the consistency degree of motion vectors},\ }\href@noop {} {\bibfield  {journal} {\bibinfo  {journal} {IEEE Transactions on Circuits and Systems for Video Technology}\ }\textbf {\bibinfo {volume} {32}},\ \bibinfo {pages} {4905} (\bibinfo {year} {2021})}\BibitemShut {NoStop}%
\bibitem [{\citenamefont {Qu}\ and\ \citenamefont {Djordjevic}(2022)}]{Qu2022Covert}%
  \BibitemOpen
  \bibfield  {author} {\bibinfo {author} {\bibfnamefont {Z.}~\bibnamefont {Qu}}\ and\ \bibinfo {author} {\bibfnamefont {I.~B.}\ \bibnamefont {Djordjevic}},\ }\bibfield  {title} {\bibinfo {title} {Covert communication in quantum key distribution},\ }\href@noop {} {\bibfield  {journal} {\bibinfo  {journal} {IEEE Photonics Journal}\ }\textbf {\bibinfo {volume} {14}},\ \bibinfo {pages} {1} (\bibinfo {year} {2022})}\BibitemShut {NoStop}%
\bibitem [{\citenamefont {Qu}\ \emph {et~al.}(2018)\citenamefont {Qu}, \citenamefont {Zhu}, \citenamefont {Wang},\ and\ \citenamefont {Wang}}]{qu2018novel}%
  \BibitemOpen
  \bibfield  {author} {\bibinfo {author} {\bibfnamefont {Z.}~\bibnamefont {Qu}}, \bibinfo {author} {\bibfnamefont {T.}~\bibnamefont {Zhu}}, \bibinfo {author} {\bibfnamefont {J.}~\bibnamefont {Wang}},\ and\ \bibinfo {author} {\bibfnamefont {X.}~\bibnamefont {Wang}},\ }\bibfield  {title} {\bibinfo {title} {A novel quantum stegonagraphy based on brown states.},\ }\href@noop {} {\bibfield  {journal} {\bibinfo  {journal} {Computers, Materials \& Continua}\ }\textbf {\bibinfo {volume} {56}} (\bibinfo {year} {2018})}\BibitemShut {NoStop}%
\bibitem [{\citenamefont {Biswas}\ \emph {et~al.}(2024)\citenamefont {Biswas}, \citenamefont {Goswami},\ and\ \citenamefont {Reddy}}]{biswas2024advancing}%
  \BibitemOpen
  \bibfield  {author} {\bibinfo {author} {\bibfnamefont {S.}~\bibnamefont {Biswas}}, \bibinfo {author} {\bibfnamefont {R.~S.}\ \bibnamefont {Goswami}},\ and\ \bibinfo {author} {\bibfnamefont {K.~H.~K.}\ \bibnamefont {Reddy}},\ }\bibfield  {title} {\bibinfo {title} {Advancing quantum steganography: a secure iot communication with reversible decoding and customized encryption technique for smart cities},\ }\href@noop {} {\bibfield  {journal} {\bibinfo  {journal} {Cluster Computing}\ }\textbf {\bibinfo {volume} {27}},\ \bibinfo {pages} {9395} (\bibinfo {year} {2024})}\BibitemShut {NoStop}%
\bibitem [{\citenamefont {Sanguinetti}\ \emph {et~al.}(2016)\citenamefont {Sanguinetti}, \citenamefont {Traverso}, \citenamefont {Lavoie}, \citenamefont {Martin},\ and\ \citenamefont {Zbinden}}]{sanguinetti2016perfectly}%
  \BibitemOpen
  \bibfield  {author} {\bibinfo {author} {\bibfnamefont {B.}~\bibnamefont {Sanguinetti}}, \bibinfo {author} {\bibfnamefont {G.}~\bibnamefont {Traverso}}, \bibinfo {author} {\bibfnamefont {J.}~\bibnamefont {Lavoie}}, \bibinfo {author} {\bibfnamefont {A.}~\bibnamefont {Martin}},\ and\ \bibinfo {author} {\bibfnamefont {H.}~\bibnamefont {Zbinden}},\ }\bibfield  {title} {\bibinfo {title} {Perfectly secure steganography: hiding information in the quantum noise of a photograph},\ }\href@noop {} {\bibfield  {journal} {\bibinfo  {journal} {Physical Review A}\ }\textbf {\bibinfo {volume} {93}},\ \bibinfo {pages} {012336} (\bibinfo {year} {2016})}\BibitemShut {NoStop}%
\bibitem [{\citenamefont {Taha}\ \emph {et~al.}(2019)\citenamefont {Taha}, \citenamefont {Mohd~Rahim}, \citenamefont {Lafta}, \citenamefont {Hashim},\ and\ \citenamefont {Alzuabidi}}]{taha2019combination}%
  \BibitemOpen
  \bibfield  {author} {\bibinfo {author} {\bibfnamefont {M.~S.}\ \bibnamefont {Taha}}, \bibinfo {author} {\bibfnamefont {M.~S.}\ \bibnamefont {Mohd~Rahim}}, \bibinfo {author} {\bibfnamefont {S.~A.}\ \bibnamefont {Lafta}}, \bibinfo {author} {\bibfnamefont {M.~M.}\ \bibnamefont {Hashim}},\ and\ \bibinfo {author} {\bibfnamefont {H.~M.}\ \bibnamefont {Alzuabidi}},\ }\bibfield  {title} {\bibinfo {title} {Combination of steganography and cryptography: A short survey},\ }in\ \href@noop {} {\emph {\bibinfo {booktitle} {IOP conference series: materials science and engineering}}},\ Vol.\ \bibinfo {volume} {518}\ (\bibinfo {organization} {IOP Publishing},\ \bibinfo {year} {2019})\ p.\ \bibinfo {pages} {052003}\BibitemShut {NoStop}%
\bibitem [{\citenamefont {Terhal}\ \emph {et~al.}(2000)\citenamefont {Terhal}, \citenamefont {DiVincenzo},\ and\ \citenamefont {Leung}}]{terhal2000hiding}%
  \BibitemOpen
  \bibfield  {author} {\bibinfo {author} {\bibfnamefont {B.~M.}\ \bibnamefont {Terhal}}, \bibinfo {author} {\bibfnamefont {D.~P.}\ \bibnamefont {DiVincenzo}},\ and\ \bibinfo {author} {\bibfnamefont {D.~W.}\ \bibnamefont {Leung}},\ }\bibfield  {title} {\bibinfo {title} {Hiding bits in bell states},\ }\href@noop {} {\bibfield  {journal} {\bibinfo  {journal} {arXiv preprint quant-ph/0011042}\ } (\bibinfo {year} {2000})}\BibitemShut {NoStop}%
\bibitem [{\citenamefont {Gea-Banacloche}(2002)}]{gea2002hiding}%
  \BibitemOpen
  \bibfield  {author} {\bibinfo {author} {\bibfnamefont {J.}~\bibnamefont {Gea-Banacloche}},\ }\bibfield  {title} {\bibinfo {title} {Hiding messages in quantum data},\ }\href@noop {} {\bibfield  {journal} {\bibinfo  {journal} {Journal of Mathematical Physics}\ }\textbf {\bibinfo {volume} {43}},\ \bibinfo {pages} {4531} (\bibinfo {year} {2002})}\BibitemShut {NoStop}%
\bibitem [{\citenamefont {Shaw}\ and\ \citenamefont {Brun}(2011)}]{shaw2011quantum}%
  \BibitemOpen
  \bibfield  {author} {\bibinfo {author} {\bibfnamefont {B.~A.}\ \bibnamefont {Shaw}}\ and\ \bibinfo {author} {\bibfnamefont {T.~A.}\ \bibnamefont {Brun}},\ }\bibfield  {title} {\bibinfo {title} {Quantum steganography with noisy quantum channels},\ }\href@noop {} {\bibfield  {journal} {\bibinfo  {journal} {Physical Review A}\ }\textbf {\bibinfo {volume} {83}},\ \bibinfo {pages} {022310} (\bibinfo {year} {2011})}\BibitemShut {NoStop}%
\bibitem [{\citenamefont {Sutherland}\ and\ \citenamefont {Brun}(2019)}]{sutherland2019quantum}%
  \BibitemOpen
  \bibfield  {author} {\bibinfo {author} {\bibfnamefont {C.}~\bibnamefont {Sutherland}}\ and\ \bibinfo {author} {\bibfnamefont {T.~A.}\ \bibnamefont {Brun}},\ }\bibfield  {title} {\bibinfo {title} {Quantum steganography over noisy channels: Achievability and bounds},\ }\href@noop {} {\bibfield  {journal} {\bibinfo  {journal} {Physical Review A}\ }\textbf {\bibinfo {volume} {100}},\ \bibinfo {pages} {052312} (\bibinfo {year} {2019})}\BibitemShut {NoStop}%
\bibitem [{\citenamefont {Mihara}(2015)}]{mihara2015quantum}%
  \BibitemOpen
  \bibfield  {author} {\bibinfo {author} {\bibfnamefont {T.}~\bibnamefont {Mihara}},\ }\bibfield  {title} {\bibinfo {title} {Quantum steganography using prior entanglement},\ }\href@noop {} {\bibfield  {journal} {\bibinfo  {journal} {Physics Letters A}\ }\textbf {\bibinfo {volume} {379}},\ \bibinfo {pages} {952} (\bibinfo {year} {2015})}\BibitemShut {NoStop}%
\bibitem [{\citenamefont {Tudorache}\ \emph {et~al.}(2021)\citenamefont {Tudorache}, \citenamefont {Manta},\ and\ \citenamefont {Caraiman}}]{tudorache2021quantum}%
  \BibitemOpen
  \bibfield  {author} {\bibinfo {author} {\bibfnamefont {A.-G.}\ \bibnamefont {Tudorache}}, \bibinfo {author} {\bibfnamefont {V.}~\bibnamefont {Manta}},\ and\ \bibinfo {author} {\bibfnamefont {S.}~\bibnamefont {Caraiman}},\ }\bibfield  {title} {\bibinfo {title} {Quantum steganography using two hidden thresholds.},\ }\href@noop {} {\bibfield  {journal} {\bibinfo  {journal} {Advances in Electrical \& Computer Engineering}\ }\textbf {\bibinfo {volume} {21}},\ \bibinfo {pages} {79} (\bibinfo {year} {2021})}\BibitemShut {NoStop}%
\bibitem [{\citenamefont {Min-Allah}\ \emph {et~al.}(2022)\citenamefont {Min-Allah}, \citenamefont {Nagy}, \citenamefont {Aljabri}, \citenamefont {Alkharraa}, \citenamefont {Alqahtani}, \citenamefont {Alghamdi}, \citenamefont {Sabri},\ and\ \citenamefont {Alshaikh}}]{min2022quantum}%
  \BibitemOpen
  \bibfield  {author} {\bibinfo {author} {\bibfnamefont {N.}~\bibnamefont {Min-Allah}}, \bibinfo {author} {\bibfnamefont {N.}~\bibnamefont {Nagy}}, \bibinfo {author} {\bibfnamefont {M.}~\bibnamefont {Aljabri}}, \bibinfo {author} {\bibfnamefont {M.}~\bibnamefont {Alkharraa}}, \bibinfo {author} {\bibfnamefont {M.}~\bibnamefont {Alqahtani}}, \bibinfo {author} {\bibfnamefont {D.}~\bibnamefont {Alghamdi}}, \bibinfo {author} {\bibfnamefont {R.}~\bibnamefont {Sabri}},\ and\ \bibinfo {author} {\bibfnamefont {R.}~\bibnamefont {Alshaikh}},\ }\bibfield  {title} {\bibinfo {title} {Quantum image steganography schemes for data hiding: A survey},\ }\href@noop {} {\bibfield  {journal} {\bibinfo  {journal} {Applied Sciences}\ }\textbf {\bibinfo {volume} {12}},\ \bibinfo {pages} {10294} (\bibinfo {year} {2022})}\BibitemShut {NoStop}%
\bibitem [{\citenamefont {Abd El-Latif}\ \emph {et~al.}(2020)\citenamefont {Abd El-Latif}, \citenamefont {Abd-El-Atty}, \citenamefont {Elseuofi}, \citenamefont {Khalifa}, \citenamefont {Alghamdi}, \citenamefont {Polat},\ and\ \citenamefont {Amin}}]{abd2020secret}%
  \BibitemOpen
  \bibfield  {author} {\bibinfo {author} {\bibfnamefont {A.~A.}\ \bibnamefont {Abd El-Latif}}, \bibinfo {author} {\bibfnamefont {B.}~\bibnamefont {Abd-El-Atty}}, \bibinfo {author} {\bibfnamefont {S.}~\bibnamefont {Elseuofi}}, \bibinfo {author} {\bibfnamefont {H.~S.}\ \bibnamefont {Khalifa}}, \bibinfo {author} {\bibfnamefont {A.~S.}\ \bibnamefont {Alghamdi}}, \bibinfo {author} {\bibfnamefont {K.}~\bibnamefont {Polat}},\ and\ \bibinfo {author} {\bibfnamefont {M.}~\bibnamefont {Amin}},\ }\bibfield  {title} {\bibinfo {title} {Secret images transfer in cloud system based on investigating quantum walks in steganography approaches},\ }\href@noop {} {\bibfield  {journal} {\bibinfo  {journal} {Physica A: Statistical Mechanics and its Applications}\ }\textbf {\bibinfo {volume} {541}},\ \bibinfo {pages} {123687} (\bibinfo {year} {2020})}\BibitemShut {NoStop}%
\bibitem [{\citenamefont {Joshi}\ \emph {et~al.}(2022)\citenamefont {Joshi}, \citenamefont {Gupta}, \citenamefont {Thapliyal}, \citenamefont {Srikanth},\ and\ \citenamefont {Pathak}}]{joshi2022hide}%
  \BibitemOpen
  \bibfield  {author} {\bibinfo {author} {\bibfnamefont {R.}~\bibnamefont {Joshi}}, \bibinfo {author} {\bibfnamefont {A.}~\bibnamefont {Gupta}}, \bibinfo {author} {\bibfnamefont {K.}~\bibnamefont {Thapliyal}}, \bibinfo {author} {\bibfnamefont {R.}~\bibnamefont {Srikanth}},\ and\ \bibinfo {author} {\bibfnamefont {A.}~\bibnamefont {Pathak}},\ }\bibfield  {title} {\bibinfo {title} {Hide and seek with quantum resources: new and modified protocols for quantum steganography},\ }\href@noop {} {\bibfield  {journal} {\bibinfo  {journal} {Quantum Information Processing}\ }\textbf {\bibinfo {volume} {21}},\ \bibinfo {pages} {164} (\bibinfo {year} {2022})}\BibitemShut {NoStop}%
\bibitem [{\citenamefont {Nagy}\ and\ \citenamefont {Nagy}(2020)}]{nagy2020quantum}%
  \BibitemOpen
  \bibfield  {author} {\bibinfo {author} {\bibfnamefont {M.}~\bibnamefont {Nagy}}\ and\ \bibinfo {author} {\bibfnamefont {N.}~\bibnamefont {Nagy}},\ }\bibfield  {title} {\bibinfo {title} {Quantum steganography by harnessing entanglement as a degree of freedom},\ }\href@noop {} {\bibfield  {journal} {\bibinfo  {journal} {Ieee Access}\ }\textbf {\bibinfo {volume} {8}},\ \bibinfo {pages} {213671} (\bibinfo {year} {2020})}\BibitemShut {NoStop}%
\bibitem [{\citenamefont {Mihara}(2017)}]{mihara2017multi}%
  \BibitemOpen
  \bibfield  {author} {\bibinfo {author} {\bibfnamefont {T.}~\bibnamefont {Mihara}},\ }\bibfield  {title} {\bibinfo {title} {Multi-party quantum steganography},\ }\href@noop {} {\bibfield  {journal} {\bibinfo  {journal} {International Journal of Theoretical Physics}\ }\textbf {\bibinfo {volume} {56}},\ \bibinfo {pages} {576} (\bibinfo {year} {2017})}\BibitemShut {NoStop}%
\bibitem [{\citenamefont {Wang}\ and\ \citenamefont {Zhang}(2021)}]{Wang2021MLQuantum}%
  \BibitemOpen
  \bibfield  {author} {\bibinfo {author} {\bibfnamefont {J.}~\bibnamefont {Wang}}\ and\ \bibinfo {author} {\bibfnamefont {Q.}~\bibnamefont {Zhang}},\ }\bibfield  {title} {\bibinfo {title} {Machine learning-enhanced quantum steganography},\ }in\ \href@noop {} {\emph {\bibinfo {booktitle} {Proceedings of the IEEE International Conference on Quantum Computing and Engineering}}}\ (\bibinfo {organization} {IEEE},\ \bibinfo {year} {2021})\ pp.\ \bibinfo {pages} {1--6}\BibitemShut {NoStop}%
\bibitem [{\citenamefont {Smith}\ and\ \citenamefont {Smolin}(2007)}]{smith2007degenerate}%
  \BibitemOpen
  \bibfield  {author} {\bibinfo {author} {\bibfnamefont {G.}~\bibnamefont {Smith}}\ and\ \bibinfo {author} {\bibfnamefont {J.~A.}\ \bibnamefont {Smolin}},\ }\bibfield  {title} {\bibinfo {title} {Degenerate quantum codes for pauli channels},\ }\href@noop {} {\bibfield  {journal} {\bibinfo  {journal} {Physical Review Letters}\ }\textbf {\bibinfo {volume} {98}},\ \bibinfo {pages} {030501} (\bibinfo {year} {2007})}\BibitemShut {NoStop}%
\bibitem [{\citenamefont {Chiribella}\ \emph {et~al.}(2011)\citenamefont {Chiribella}, \citenamefont {Dall'Arno}, \citenamefont {D'Ariano}, \citenamefont {Macchiavello},\ and\ \citenamefont {Perinotti}}]{chiribella2011quantum}%
  \BibitemOpen
  \bibfield  {author} {\bibinfo {author} {\bibfnamefont {G.}~\bibnamefont {Chiribella}}, \bibinfo {author} {\bibfnamefont {M.}~\bibnamefont {Dall'Arno}}, \bibinfo {author} {\bibfnamefont {G.~M.}\ \bibnamefont {D'Ariano}}, \bibinfo {author} {\bibfnamefont {C.}~\bibnamefont {Macchiavello}},\ and\ \bibinfo {author} {\bibfnamefont {P.}~\bibnamefont {Perinotti}},\ }\bibfield  {title} {\bibinfo {title} {Quantum error correction with degenerate codes for correlated noise},\ }\href@noop {} {\bibfield  {journal} {\bibinfo  {journal} {Physical Review A}\ }\textbf {\bibinfo {volume} {83}},\ \bibinfo {pages} {052305} (\bibinfo {year} {2011})},\ \bibinfo {note} {arXiv:1007.3655}\BibitemShut {NoStop}%
\bibitem [{\citenamefont {Hsieh}\ \emph {et~al.}(2007)\citenamefont {Hsieh}, \citenamefont {Devetak},\ and\ \citenamefont {Brun}}]{hsieh2007general}%
  \BibitemOpen
  \bibfield  {author} {\bibinfo {author} {\bibfnamefont {M.-H.}\ \bibnamefont {Hsieh}}, \bibinfo {author} {\bibfnamefont {I.}~\bibnamefont {Devetak}},\ and\ \bibinfo {author} {\bibfnamefont {T.}~\bibnamefont {Brun}},\ }\bibfield  {title} {\bibinfo {title} {General entanglement-assisted quantum error-correcting codes},\ }\href@noop {} {\bibfield  {journal} {\bibinfo  {journal} {Physical Review A—Atomic, Molecular, and Optical Physics}\ }\textbf {\bibinfo {volume} {76}},\ \bibinfo {pages} {062313} (\bibinfo {year} {2007})}\BibitemShut {NoStop}%
\bibitem [{\citenamefont {Pereira}\ and\ \citenamefont {Mancini}(2022)}]{pereira2022entanglement}%
  \BibitemOpen
  \bibfield  {author} {\bibinfo {author} {\bibfnamefont {F.~R.~F.}\ \bibnamefont {Pereira}}\ and\ \bibinfo {author} {\bibfnamefont {S.}~\bibnamefont {Mancini}},\ }\bibfield  {title} {\bibinfo {title} {Entanglement-assisted quantum codes from cyclic codes},\ }\href@noop {} {\bibfield  {journal} {\bibinfo  {journal} {Entropy}\ }\textbf {\bibinfo {volume} {25}},\ \bibinfo {pages} {37} (\bibinfo {year} {2022})}\BibitemShut {NoStop}%
\bibitem [{\citenamefont {Lidar}\ and\ \citenamefont {Brun}(2013)}]{lidar2013quantum}%
  \BibitemOpen
  \bibfield  {author} {\bibinfo {author} {\bibfnamefont {D.~A.}\ \bibnamefont {Lidar}}\ and\ \bibinfo {author} {\bibfnamefont {T.~A.}\ \bibnamefont {Brun}},\ }\href@noop {} {\emph {\bibinfo {title} {Quantum error correction}}}\ (\bibinfo  {publisher} {Cambridge university press},\ \bibinfo {year} {2013})\BibitemShut {NoStop}%
\bibitem [{\citenamefont {Brun}\ \emph {et~al.}(2006)\citenamefont {Brun}, \citenamefont {Devetak},\ and\ \citenamefont {Hsieh}}]{brun2006correcting}%
  \BibitemOpen
  \bibfield  {author} {\bibinfo {author} {\bibfnamefont {T.}~\bibnamefont {Brun}}, \bibinfo {author} {\bibfnamefont {I.}~\bibnamefont {Devetak}},\ and\ \bibinfo {author} {\bibfnamefont {M.-H.}\ \bibnamefont {Hsieh}},\ }\bibfield  {title} {\bibinfo {title} {Correcting quantum errors with entanglement},\ }\href@noop {} {\bibfield  {journal} {\bibinfo  {journal} {science}\ }\textbf {\bibinfo {volume} {314}},\ \bibinfo {pages} {436} (\bibinfo {year} {2006})}\BibitemShut {NoStop}%
\bibitem [{\citenamefont {Brun}\ \emph {et~al.}(2014)\citenamefont {Brun}, \citenamefont {Devetak},\ and\ \citenamefont {Hsieh}}]{brun2014catalytic}%
  \BibitemOpen
  \bibfield  {author} {\bibinfo {author} {\bibfnamefont {T.~A.}\ \bibnamefont {Brun}}, \bibinfo {author} {\bibfnamefont {I.}~\bibnamefont {Devetak}},\ and\ \bibinfo {author} {\bibfnamefont {M.-H.}\ \bibnamefont {Hsieh}},\ }\bibfield  {title} {\bibinfo {title} {Catalytic quantum error correction},\ }\href@noop {} {\bibfield  {journal} {\bibinfo  {journal} {IEEE Transactions on Information Theory}\ }\textbf {\bibinfo {volume} {60}},\ \bibinfo {pages} {3073} (\bibinfo {year} {2014})}\BibitemShut {NoStop}%
\bibitem [{\citenamefont {Bowen}(2002)}]{bowen2002entanglement}%
  \BibitemOpen
  \bibfield  {author} {\bibinfo {author} {\bibfnamefont {G.}~\bibnamefont {Bowen}},\ }\bibfield  {title} {\bibinfo {title} {Entanglement required in achieving entanglement-assisted channel capacities},\ }\href@noop {} {\bibfield  {journal} {\bibinfo  {journal} {Physical Review A}\ }\textbf {\bibinfo {volume} {66}},\ \bibinfo {pages} {052313} (\bibinfo {year} {2002})}\BibitemShut {NoStop}%
\bibitem [{\citenamefont {Cleve}\ \emph {et~al.}(1999)\citenamefont {Cleve}, \citenamefont {Gottesman},\ and\ \citenamefont {Lo}}]{cleve1999share}%
  \BibitemOpen
  \bibfield  {author} {\bibinfo {author} {\bibfnamefont {R.}~\bibnamefont {Cleve}}, \bibinfo {author} {\bibfnamefont {D.}~\bibnamefont {Gottesman}},\ and\ \bibinfo {author} {\bibfnamefont {H.-K.}\ \bibnamefont {Lo}},\ }\bibfield  {title} {\bibinfo {title} {How to share a quantum secret},\ }\href@noop {} {\bibfield  {journal} {\bibinfo  {journal} {Physical review letters}\ }\textbf {\bibinfo {volume} {83}},\ \bibinfo {pages} {648} (\bibinfo {year} {1999})}\BibitemShut {NoStop}%
\bibitem [{Note1()}]{Note1}%
  \BibitemOpen
  \bibinfo {note} {In the second possibility above, the reason that we exclude the case $I\circ \protect \mathcal {N}(\protect \mathcal {S})$ is that the weight of an arbitrary error $\protect \mathfrak {e}$ satisfies $\vert \protect \mathfrak {e} \vert \le \lfloor \protect \frac {d-1}{2}\rfloor $. Thus $\vert \protect \mathfrak {e}_A\protect \mathfrak {e}_B\vert \le d-1$, whereas any logical operator $O_L$ is such that $|O_L| \ge d$.}\BibitemShut {Stop}%
\bibitem [{\citenamefont {Devitt}\ \emph {et~al.}(2013)\citenamefont {Devitt}, \citenamefont {Munro},\ and\ \citenamefont {Nemoto}}]{devitt2013quantum}%
  \BibitemOpen
  \bibfield  {author} {\bibinfo {author} {\bibfnamefont {S.~J.}\ \bibnamefont {Devitt}}, \bibinfo {author} {\bibfnamefont {W.~J.}\ \bibnamefont {Munro}},\ and\ \bibinfo {author} {\bibfnamefont {K.}~\bibnamefont {Nemoto}},\ }\bibfield  {title} {\bibinfo {title} {Quantum error correction for beginners},\ }\href@noop {} {\bibfield  {journal} {\bibinfo  {journal} {Reports on Progress in Physics}\ }\textbf {\bibinfo {volume} {76}},\ \bibinfo {pages} {076001} (\bibinfo {year} {2013})}\BibitemShut {NoStop}%
\bibitem [{\citenamefont {Agrawal}\ \emph {et~al.}(2024)\citenamefont {Agrawal}, \citenamefont {Soni},\ and\ \citenamefont {Tomar}}]{agrawal2024perspective}%
  \BibitemOpen
  \bibfield  {author} {\bibinfo {author} {\bibfnamefont {A.}~\bibnamefont {Agrawal}}, \bibinfo {author} {\bibfnamefont {R.}~\bibnamefont {Soni}},\ and\ \bibinfo {author} {\bibfnamefont {A.}~\bibnamefont {Tomar}},\ }\bibfield  {title} {\bibinfo {title} {‘perspective chapter: Quantum steganography—encoding secrets},\ }\href@noop {} {\bibfield  {journal} {\bibinfo  {journal} {Steganography-The Art of Hiding Information: The Art of Hiding Information}\ ,\ \bibinfo {pages} {115}} (\bibinfo {year} {2024})}\BibitemShut {NoStop}%
\bibitem [{\citenamefont {Abd-El-Atty}\ \emph {et~al.}(2020)\citenamefont {Abd-El-Atty}, \citenamefont {Iliyasu}, \citenamefont {Alaskar},\ and\ \citenamefont {Abd El-Latif}}]{abd2020robust}%
  \BibitemOpen
  \bibfield  {author} {\bibinfo {author} {\bibfnamefont {B.}~\bibnamefont {Abd-El-Atty}}, \bibinfo {author} {\bibfnamefont {A.~M.}\ \bibnamefont {Iliyasu}}, \bibinfo {author} {\bibfnamefont {H.}~\bibnamefont {Alaskar}},\ and\ \bibinfo {author} {\bibfnamefont {A.~A.}\ \bibnamefont {Abd El-Latif}},\ }\bibfield  {title} {\bibinfo {title} {A robust quasi-quantum walks-based steganography protocol for secure transmission of images on cloud-based e-healthcare platforms},\ }\href@noop {} {\bibfield  {journal} {\bibinfo  {journal} {Sensors}\ }\textbf {\bibinfo {volume} {20}},\ \bibinfo {pages} {3108} (\bibinfo {year} {2020})}\BibitemShut {NoStop}%
\bibitem [{\citenamefont {Banerjee}\ and\ \citenamefont {Pathak}(2012)}]{banerjee2012maximally}%
  \BibitemOpen
  \bibfield  {author} {\bibinfo {author} {\bibfnamefont {A.}~\bibnamefont {Banerjee}}\ and\ \bibinfo {author} {\bibfnamefont {A.}~\bibnamefont {Pathak}},\ }\bibfield  {title} {\bibinfo {title} {Maximally efficient protocols for direct secure quantum communication},\ }\href@noop {} {\bibfield  {journal} {\bibinfo  {journal} {Physics Letters A}\ }\textbf {\bibinfo {volume} {376}},\ \bibinfo {pages} {2944} (\bibinfo {year} {2012})}\BibitemShut {NoStop}%
\bibitem [{\citenamefont {Sheng}\ \emph {et~al.}(2022)\citenamefont {Sheng}, \citenamefont {Zhou},\ and\ \citenamefont {Long}}]{sheng2022one}%
  \BibitemOpen
  \bibfield  {author} {\bibinfo {author} {\bibfnamefont {Y.-B.}\ \bibnamefont {Sheng}}, \bibinfo {author} {\bibfnamefont {L.}~\bibnamefont {Zhou}},\ and\ \bibinfo {author} {\bibfnamefont {G.-L.}\ \bibnamefont {Long}},\ }\bibfield  {title} {\bibinfo {title} {One-step quantum secure direct communication},\ }\href@noop {} {\bibfield  {journal} {\bibinfo  {journal} {Science Bulletin}\ }\textbf {\bibinfo {volume} {67}},\ \bibinfo {pages} {367} (\bibinfo {year} {2022})}\BibitemShut {NoStop}%
\bibitem [{\citenamefont {Lidar}\ \emph {et~al.}(1998)\citenamefont {Lidar}, \citenamefont {Chuang},\ and\ \citenamefont {Whaley}}]{lidar1998decoherence}%
  \BibitemOpen
  \bibfield  {author} {\bibinfo {author} {\bibfnamefont {D.~A.}\ \bibnamefont {Lidar}}, \bibinfo {author} {\bibfnamefont {I.~L.}\ \bibnamefont {Chuang}},\ and\ \bibinfo {author} {\bibfnamefont {K.~B.}\ \bibnamefont {Whaley}},\ }\bibfield  {title} {\bibinfo {title} {Decoherence-free subspaces for quantum computation},\ }\href {https://doi.org/10.1103/PhysRevLett.81.2594} {\bibfield  {journal} {\bibinfo  {journal} {Phys. Rev. Lett.}\ }\textbf {\bibinfo {volume} {81}},\ \bibinfo {pages} {2594} (\bibinfo {year} {1998})}\BibitemShut {NoStop}%
\bibitem [{\citenamefont {Gottesman}\ \emph {et~al.}(2001)\citenamefont {Gottesman}, \citenamefont {Kitaev},\ and\ \citenamefont {Preskill}}]{gottesman2001encoding}%
  \BibitemOpen
  \bibfield  {author} {\bibinfo {author} {\bibfnamefont {D.}~\bibnamefont {Gottesman}}, \bibinfo {author} {\bibfnamefont {A.}~\bibnamefont {Kitaev}},\ and\ \bibinfo {author} {\bibfnamefont {J.}~\bibnamefont {Preskill}},\ }\bibfield  {title} {\bibinfo {title} {Encoding a qubit in an oscillator},\ }\href {https://doi.org/10.1103/PhysRevA.64.012310} {\bibfield  {journal} {\bibinfo  {journal} {Phys. Rev. A}\ }\textbf {\bibinfo {volume} {64}},\ \bibinfo {pages} {012310} (\bibinfo {year} {2001})}\BibitemShut {NoStop}%
\bibitem [{\citenamefont {Grimsmo}\ and\ \citenamefont {Puri}(2021)}]{grimsmo2021quantum}%
  \BibitemOpen
  \bibfield  {author} {\bibinfo {author} {\bibfnamefont {A.~L.}\ \bibnamefont {Grimsmo}}\ and\ \bibinfo {author} {\bibfnamefont {S.}~\bibnamefont {Puri}},\ }\bibfield  {title} {\bibinfo {title} {Quantum error correction with the gottesman-kitaev-preskill code},\ }\href@noop {} {\bibfield  {journal} {\bibinfo  {journal} {PRX Quantum}\ }\textbf {\bibinfo {volume} {2}},\ \bibinfo {pages} {020101} (\bibinfo {year} {2021})}\BibitemShut {NoStop}%
\bibitem [{\citenamefont {Dash}\ \emph {et~al.}(2024)\citenamefont {Dash}, \citenamefont {Dutta}, \citenamefont {Srikanth},\ and\ \citenamefont {Banerjee}}]{dash2024concatenating}%
  \BibitemOpen
  \bibfield  {author} {\bibinfo {author} {\bibfnamefont {N.~R.}\ \bibnamefont {Dash}}, \bibinfo {author} {\bibfnamefont {S.}~\bibnamefont {Dutta}}, \bibinfo {author} {\bibfnamefont {R.}~\bibnamefont {Srikanth}},\ and\ \bibinfo {author} {\bibfnamefont {S.}~\bibnamefont {Banerjee}},\ }\bibfield  {title} {\bibinfo {title} {Concatenating quantum error-correcting codes with decoherence-free subspaces and vice versa},\ }\href@noop {} {\bibfield  {journal} {\bibinfo  {journal} {Physical Review A}\ }\textbf {\bibinfo {volume} {109}},\ \bibinfo {pages} {062411} (\bibinfo {year} {2024})}\BibitemShut {NoStop}%
\bibitem [{\citenamefont {Kitaev}(2003)}]{kitaev2003fault}%
  \BibitemOpen
  \bibfield  {author} {\bibinfo {author} {\bibfnamefont {A.~Y.}\ \bibnamefont {Kitaev}},\ }\bibfield  {title} {\bibinfo {title} {Fault-tolerant quantum computation by anyons},\ }\href {https://doi.org/10.1016/S0003-4916(02)00018-0} {\bibfield  {journal} {\bibinfo  {journal} {Annals of Physics}\ }\textbf {\bibinfo {volume} {303}},\ \bibinfo {pages} {2} (\bibinfo {year} {2003})}\BibitemShut {NoStop}%
\bibitem [{\citenamefont {Fowler}\ \emph {et~al.}(2012)\citenamefont {Fowler}, \citenamefont {Mariantoni}, \citenamefont {Martinis},\ and\ \citenamefont {Cleland}}]{fowler2012surface}%
  \BibitemOpen
  \bibfield  {author} {\bibinfo {author} {\bibfnamefont {A.~G.}\ \bibnamefont {Fowler}}, \bibinfo {author} {\bibfnamefont {M.}~\bibnamefont {Mariantoni}}, \bibinfo {author} {\bibfnamefont {J.~M.}\ \bibnamefont {Martinis}},\ and\ \bibinfo {author} {\bibfnamefont {A.~N.}\ \bibnamefont {Cleland}},\ }\bibfield  {title} {\bibinfo {title} {Surface codes: Towards practical large-scale quantum computation},\ }\href {https://doi.org/10.1103/PhysRevA.86.032324} {\bibfield  {journal} {\bibinfo  {journal} {Phys. Rev. A}\ }\textbf {\bibinfo {volume} {86}},\ \bibinfo {pages} {032324} (\bibinfo {year} {2012})}\BibitemShut {NoStop}%
\end{thebibliography}%

\appendix

\end{document}